\documentclass[final, 3p, times]{elsarticle}
\usepackage{hyperref}
\usepackage{amsmath,amssymb,amsfonts}
\usepackage{algorithmic}
\usepackage{textcomp}
\usepackage{amsthm}
\usepackage{algorithm}
\usepackage{multirow}
\usepackage{subcaption}

\newtheorem{thm}{Theorem}
\newtheorem{lem}[thm]{Lemma}
\newtheorem{proposition}{Proposition}
\newtheorem{rmk}{Remark}
\newtheorem{assumption}{Assumption}
\newtheorem{defn}{Definition}
\newtheorem{problem}{Problem}

\usepackage{balance}
\usepackage{tikzscale}
\usepackage{pgfplots}
\usepackage{tikz}
\usepackage{xcolor}
\usepackage{graphicx}
\usepackage{url}
\usepackage{bbm}

\DeclareMathOperator*{\esssup}{ess\,sup}
\usepackage{mathtools,amssymb,lipsum}

\usepackage{cuted}
\usepackage{ulem}

\journal{Aritificial Intelligence Journal }
\begin{document}

\begin{frontmatter}

\title{Risk-Averse Receding Horizon Motion Planning \\ for Obstacle Avoidance using Coherent Risk Measures}
\author[mymainaddress]{Anushri Dixit\corref{mycorrespondingauthor}}
\cortext[mycorrespondingauthor]{Corresponding author}
\ead{adixit@caltech.edu}

\author[mymainaddress]{Mohamadreza Ahmadi}
\ead{mrahmadi@caltech.edu}
\author[mymainaddress]{Joel W. Burdick}
\ead{jwb@robotics.caltech.edu}

\address[mymainaddress]{1200 E California Blvd, Pasadena, CA 91125}
\begin{abstract}
This paper studies the problem of risk-averse receding horizon motion planning for agents with uncertain dynamics, in the presence of stochastic, dynamic obstacles. We propose a model predictive control (MPC) scheme that formulates the obstacle avoidance constraint using coherent risk measures. To handle disturbances, or process noise, in the state dynamics, the state constraints are tightened in a risk-aware manner to provide a disturbance feedback policy. We also propose a waypoint following algorithm that uses the proposed MPC scheme for discrete distributions and prove its risk-sensitive recursive feasibility while guaranteeing finite-time task completion. We further investigate some commonly used coherent risk metrics, namely, conditional value-at-risk (CVaR), entropic value-at-risk (EVaR), and g-entropic risk measures, and propose a tractable incorporation within MPC. We illustrate our framework via simulation studies. 
\end{abstract}


\end{frontmatter}


\section{Introduction}
Autonomous robots must increasingly plan motions in unstructured and uncertain environments with safety guarantees. Some applications where safe planning is required include autonomous traversal over extreme terrain in GPS-denied subterranean environments~\cite{bouman2020autonomous, fan2021step}, inspection of planetary environments~\cite{daftry2022mlnav}, search and rescue missions caused by natural disasters~\cite{nagatani2013emergency,seraj2020coordinated}, and autonomous driving~\cite{rosolia2017lmpc}. These applications present challenges at all the levels of planning and control~\cite{rosolia2020unified}. The lowest control level requires a good physical model for accurate motion prediction. To ensure robustness and safety, these models are often equipped with low-level controllers that leverage tools from robust control and invariant set theory~\cite{ames2017cbf}. At the intermediate level, algorithms must plan paths that are dynamically feasible, obstacle-free, and account for uncertainty in the motion dynamics, sensor measurements, and the environment. Several existing algorithms (model predictive control and A*-based graph planners to name a few) tackle some or all of these issues~\cite{blackmore2011CC, jasourrisk}. \textcolor{black}{Sampling-based planners like CC-RRT~\cite{luders2010chance, Aoude2013ProbabilisticallySM} are another popular way to obtain dynamically feasible trajectories that satisfy constraints, they however do not guarantee any form of optimality. Other techniques use RRT-based techniques to compute reachable sets for solving an approximate stochastic optimal control problem~\cite{lindemann2021robust}. } At the highest level, robots must reason about their (uncertain) environment and decide on what tasks to do. Partially Observable Markov Decision Processes (POMDPs) are popular models for such sequential planning tasks~\cite{kim2021plgrim}. Our work looks at the problem of obstacle avoidance using model predictive control (MPC) techniques.

MPC is widely used for robotic motion planning because it incorporates robot dynamics and state and control constraints in a receding horizon fashion~\cite{borrelli2003constrained,nair2021stochastic}. There are many ways to incorporate uncertainty in MPC. Robust MPC accounts for worst-case disturbances in a set of bounded uncertainties~\cite{bemporad1999robust}. This approach is often too conservative, since it does not account for the distribution of the uncertainties. Stochastic MPC~\cite{mesbah2016SMPC} minimizes the expected value of a cost function, while respecting a bound on the probability of violating the state and control constraints.The chance constraints in stochastic MPC do not usefully account for events in the tail of the uncertainty distribution, and the policy that results from an expected cost function minimizes the cost on average. In this work, we optimize for policies that have risk-averse behavior: the policies are not as conservative as in the robust case but account for ``risky'' outcomes in the tail of the uncertainty distribution and therefore perform better in practice.

There are many ways to incorporate risk into a control strategy~\cite{jasour2019risk}, such as chance constraints~\cite{ono2015chance,han2022non}, \textcolor{black}{exponential utility functions~\cite{Hyeon2020fast, moehle2021risk, koenig1994risk}}, and \textcolor{black}{distributional robustness}~\cite{xu2010distributionally,chen2022distributionally},\textcolor{black}{~\cite{RENGANATHAN202015530,nair2022collision}}. However, applications in autonomy and robotics require more ``nuanced assessments of risk''~\cite{majumdar2020should}. Artzner \textit{et. al.}~\cite{artzner1999coherent} characterized a set of \textit{coherent risk measures} that have natural and desirable properties. This paper focuses on measures which are widely used in finance and operations research, among other fields.
 
Motion planning based on coherent risk measures has previously been considered.  In~\cite{singh2018framework}, the authors provided an MPC scheme for a discrete-time, \textcolor{black}{linear} dynamical system with process noise whose objective was a Conditional Value-at-Risk (CVaR) measure. They further provided Lyapunov conditions for risk-sensitive exponential stability. \textcolor{black}{In~\cite{wang2021adaptive}, the authors consider a stochastic search algorithm for CVaR cost-based optimization for uncertain, nonlinear systems. }In~\cite{hakobyan2019cvar}, the authors devised an MPC scheme to avoid randomly moving obstacles using a CVaR risk metric. Similar results were obtained in~\cite{dixit2020risksensitive} on Entropic Value-at-Risk (EVaR) metric for obstacle avoidance with additional guarantees of recursive feasibility and finite-time task completion while following a set of waypoints. Risk-sensitive obstacle avoidance has also been tackled through CVaR-based control barrier functions \textcolor{black}{for nonlinear systems} in~\cite{ahmadi2020cvar} with application to bipedal robot locomotion. In~\cite{Sopasakis2019}, the authors considered multistage risk-averse and risk-constrained optimal control for general coherent risk measures with conic representations. A scenario tree-based branch MPC framework with feedback policies that account for a tradeoff between robustness and performance through CVaR metrics was proposed in~\cite{chen2022interactive}. \textcolor{black}{A learning-based distributionally-robust CVaR formulation was considered for adaptive cruise control applications in~\cite{SCHUURMANS202015128}.}
 
This paper provides a framework for risk-averse model predictive control with obstacle avoidance constraints.  This work is an extension of previous work~\cite{dixit2020risksensitive} that allowed for randomly moving obstacles while using entropic value-at-risk as the risk metric. This paper allows for a linear discrete-time system to be affected by both process noise as well as measurement noise in the sensing of obstacle position and orientation. The control input is parameterized as a disturbance feedback policy \textcolor{black}{as opposed to optimizing for open-loop  control actions that are more conservative}. Additionally, the MPC scheme in this work allows for a general class of coherent risk measures \textcolor{black}{and for arbitrary uncertainty distributions}. Coherent risk measures can be expressed as a distributionally-robust expectation, i.e, the risk is equivalently expressed as the worst-case expectation over a convex, closed set of distributions. We use this property of distributional robustness extensively throughout this paper. We reformulate the risk-aware MPC with obstacle avoidance constraints as a convex, mixed-integer program. \textcolor{black}{We further provide constraint tightening techniques that reduce the problem complexity from having exponential growth (with horizon length) of the number of constraints to depending polynomially on the horizon length. Hence, we provide a general risk-aware MPC framework for dynamic obstacle avoidance in the presence of state and measurement noise that allows for a large class of coherent risk measures (including CVaR, EVaR, total variation distance and other f-divergence based risk metrics). We compute a feedback policy that enjoys recursive feasibility and finite-time task completion guarantees in probability. We discuss the tractability of this approach for various commonly used coherent risk measures through numerical simulations compare it against standard stochastic MPC frameworks.}

This paper is organized as follows. First, we review relevant facts on coherent risk measures and some commonly used examples of these measures in Section~\ref{section:prelim}. Section~\ref{sec:problem_state} presents the main problem studied in this paper. Section~\ref{sec:mpc_form} proposes a problem reformulation based on convex mixed integer programming to solve the risk-averse receding horizon path planning problem. It also discusses the properties of feasibility and finite-time task completion with some confidence. Section~\ref{sec:results} illustrates the method via numerical experiments. We conclude with a discussion of our contributions and avenues of future work in Section~\ref{sec:conclusions}. The Appendix contains the proofs not presented in the preceding sections.

\textbf{Notation: } We denote by $\mathbb{R}^n$ the $n$-dimensional Euclidean space, $\mathbb{R}_{\ge 0}$ the non-negative reals, and $\mathbb{Z}_{\ge0}$ the set of non-negative integers. The index set, $\{k, k+1, \dotsc, k+N \}$ is denoted by $\mathbb{Z}_k^{k+N}$. Throughout the paper, a bold font denotes a vector and $(\cdot)^\top$ is its transpose, \textit{e.g.,} $\boldsymbol{a} = (a_1, \ldots, a_n)^\top$, with $n\in \{1,2,\ldots\}$. For vector $\boldsymbol{a}$, we use $\boldsymbol{a}\succeq (\preceq) \boldsymbol{0}$ to denote element-wise non-negativity (non-positivity), $\boldsymbol{a}\equiv \boldsymbol{0}$ to show all elements of $\boldsymbol{a}$ are zero, and $|\boldsymbol{a}|$ to denote the element-wise absolute value of $\boldsymbol{a}$. For vectors $a,b \in \mathbb{R}^n$, we denote their inner product by $\langle \boldsymbol{a}, \boldsymbol{b} \rangle$, \textit{i.e.,} $\langle \boldsymbol{a}, \boldsymbol{b} \rangle=\boldsymbol{a}^\top \boldsymbol{b}$. For a finite set $\mathcal{A}$, denote its power set $2^\mathcal{A}$.
For  a probability space $(\Omega, \mathcal{F}, \mathbb{P})$ and a constant $p \in [1,\infty)$, $\mathcal{L}_p(\Omega, \mathcal{F}, \mathbb{P})$ denotes the vector space of real valued random variables $X$ for which $\mathbb{E}|X|^p < \infty$. For probability density functions $P(X)$ and $Q(X)$, $P \ll Q$ implies that $P$ is absolutely continuous with respect to $Q$, \textit{i.e.,} if $Q(X)=0$, then $P(X)=0$.

We follow the following convention for indices that appear most commonly as sub/superscripts in this paper:
\begin{itemize}
    \item $k$ indicates the state at time $t+k$, $\boldsymbol{x}_k = \boldsymbol{x}(t+k|t)$,
    \item $j$ denotes the values associated with the $j^{\text{th}}$ possible occurrence of the random variable (from the sample space),
    \item $l, i$ represent the $l^{\text{th}}$ obstacle and the $i^{\text{th}}$ edge of the obstacle respectively.
\end{itemize}

\section{Preliminaries}\label{section:prelim}

We consider a probability space $(\Omega, \mathcal{F}, {P})$, where $\Omega$, $\mathcal{F}$, and ${P}$ are the sample space, $\sigma$-algebra over $\Omega$, and probability measure over $\mathcal{F}$ respectively. In this paper, a random variable $X: \Omega\xrightarrow[]{}\mathbb{R}$ denotes the cost of each outcome. The set of all cost random variables defined on $\Omega$ is given by $\mathfrak{F}$. A risk measure is a function that maps a cost random variable to a real number, $\rho:\mathfrak{F}\xrightarrow[]{}\mathbb{R}$. 

For constrained stochastic optimization programs, chance constraints can be reformulated using a commonly used risk measure called the \textit{Value-at-Risk} (VaR). For a given confidence level $\alpha \in (0,1)$, $\mathrm{VaR}_{1-\alpha}$ denotes the $({1-\alpha})$-quantile value of the cost variable $X$ and is defined as, 
\begin{align*}
    \text{VaR}_{1-\alpha}(X) := \inf \{ z \,|\, \mathbb{P}(X \leq z) \geq \alpha\}.
\end{align*}
It follows that $
    \text{VaR}_{1-\alpha}(X)\leq 0 \implies \mathbb{P}(X \leq 0)\geq\alpha.$
However, VaR is generally nonconvex and hard to compute. We now introduce convex and monotonic risk measures. In particular, we are interested in coherent risk measures~\cite{artzner1999coherent} that satisfy the following properties.

\begin{defn}[Coherent Risk Measures]{Consider two random variables, $X$, $X'\in\mathfrak{F}$. A coherent risk measure, $\rho:\mathfrak{F}\xrightarrow[]{}\mathbb{R}$, is a risk measure that satisfies the following properties:
\begin{enumerate}
    \item \textbf{Monotonicity} $X \leq X' \implies \rho(X) \leq \rho(X')$,
    \item \textbf{Translational invariance} $\rho(X + c) = \rho(X) + c, \, \forall c \in \mathbb{R}$,
    \item \textbf{Positive homogeneity} $\rho(\alpha X) = \alpha\rho(X), \, \forall \alpha \geq 0$, 
    \item \textbf{Subadditivity} $\rho (X + X') \leq \rho(X) + \rho(X')$.
\end{enumerate}}
\end{defn}

Another nice property of coherent risk measures is that they can be written as the worst-case expectation over a convex, bounded, and closed set of probability mass (or density) functions (pdf/pmf). This is the dual representation of a risk measure, and this set is referred to as the \textit{risk envelope}. 

\begin{defn}[Representation Theorem~\cite{artzner1999coherent}]{Every coherent risk measure can be represented in its dual form as, 
\begin{align*}
    \rho(X) = \sup_{Q \in \mathcal{Q}}\mathbb{E}_Q(X),
\end{align*}
where \textcolor{black}{there exists a family of probability measures, } $\mathcal{Q} \subset \{Q \ll {P}\}$ that is convex and closed (called the risk envelope).
} 
\end{defn}

While coherent risk measures act on a one-dimensional cost random variable, in this paper, we write $\rho(\boldsymbol{X})$, where $\boldsymbol{X}$ is a vector of cost random variables of length $n$, to mean $\rho(\boldsymbol{X}) = \begin{bmatrix}\rho(X_1), \dotsc, \rho(X_n) \end{bmatrix}^T$.

Note that VaR is generally not a coherent risk measure.  We next review some examples of coherent risk measures and their dual representation. We will apply our results to these examples.

\subsection{Conditional Value-at-Risk}

For a given confidence level $\alpha \in (0,1)$, value-at-risk $\mathrm{VaR}_{1-\alpha}$ denotes the $({1-\alpha})$-quantile value of the cost variable $X \in \mathcal{L}_p(\Omega, \mathcal{F}, \mathbb{P})$. The conditional value-at-risk $\mathrm{CVaR}_{1-\alpha}$ measures the expected loss in the $({1-\alpha})$-tail given that the threshold $\mathrm{VaR}_{1-\alpha}$ has been crossed. $\mathrm{CVaR}_{1-\alpha}$ is found as
\begin{equation}\label{eq:cvar_def}
\begin{aligned}
    \mathrm{CVaR}_{1-\alpha}(X):=&\inf_{z \in \mathbb{R}}\mathbb{E}\Bigg[z + \frac{(X-z)^{+}}{1-\alpha}\Bigg], 
\end{aligned}
\end{equation}
 where $(\cdot)^{+}=\max\{\cdot, 0\}$. A value of $\alpha \simeq 0$ corresponds to a risk-neutral case; whereas, a value of $\alpha \to 1$ is rather a risk-averse case.  
 $\mathcal{Q}$ is the risk envelope defined by,

\begin{equation}
    \mathcal{Q}:= \Big\{Q\ll P\, |\, 0\leq \frac{dQ}{dP} \leq\frac{1}{1-\alpha}\Big\},
\end{equation}
where $\frac{dQ}{dP}$ is called the Radon–Nikodym derivative and it gives the rate of change of density of one density function, $Q$, w.r.t the other, $P$. Similarly, for a discrete random variable $X \in \{x_1, x_2,\dotsc, x_J\}$ with pmf given by $p = [p(1), p(2), \dotsc, p(J)]^T$, where $p(j) = \mathbb{P}(X = x_j), j \in \mathbb{Z}_1^J$, the risk envelope translates to
\begin{equation}
    \mathcal{Q}:= \Big\{q \in \Delta_J \,|\,0\leq q(j) \leq\frac{p(j)}{1-\alpha}\, \forall j\in\{1,\dots,J\}\Big\}
\end{equation}
where $\Delta_J$ is the probability simplex, $ \Delta_J:= \{q \in \mathbb{R}^J\,|\, q \geq 0, \,\sum_{j=1}^{J}q(j) = 1\}$.

\subsection{Entropic Value-at-Risk}
EVaR, derived using the Chernoff inequality for VaR, is the tightest upper bound for VaR and CVaR. The $\mathrm{EVaR}_{1-\alpha}$ of random variable $X$ is given~by
\begin{equation} \label{eq:evar}
   \mathrm{EVaR}_{1-\alpha}(X):= \inf_{z > 0 }\Bigg[z^{-1}\ln\frac{\mathbb{E}[e^{Xz}]}{1-\alpha}\Bigg].
\end{equation}
%
Similar to $\mathrm{CVaR}_{1-\alpha}$, for $\mathrm{EVaR}_{1-\alpha}$, the limit $\alpha \to 0$ corresponds to a risk-neutral case; whereas, $\alpha \to 1$ corresponds to a risk-averse case. In fact, it was demonstrated in~\cite[Proposition 3.2]{ahmadi2012entropic} that $\lim_{{\alpha}\to 1} \mathrm{EVaR}_{{1-\alpha}}(X) = \esssup(X)$, where $\esssup(X)$ is the worst case value of $X$. 

For EVaR, the risk envelope $\mathcal{Q}$ for a continuous random variable with the pdf $P$ is defined as the epigraph of the KL divergence,
\begin{equation}
  { \mathcal{Q}:= \Big\{Q\ll P\, | \, D_{KL}(Q||P):=\int\frac{dQ}{dP}\big(\ln\frac{dQ}{dP}\big)dP \leq -\ln(1-\alpha)\Big\},}
\end{equation}
\noindent where $D_{KL}(Q||P)$ denotes the KL divergence between the distributions $Q$ and $P$. For some $x,y \in \mathbb{R}$, $D_{KL}(x||y)$ can be written in the form of the exponential cone, $K_{exp}$:
\begin{equation*}
    t\geq x\ln(x/y) \iff  (y, x, -t) \in K_{exp}.
\end{equation*}
Similarly, for a discrete random variable $X \in \{x_1, x_2,\dotsc, x_J\}$ with pmf given by $p = [p(1), p(2), \dotsc, p(J)]^T$, where $p(j) = \mathbb{P}(X = x_j), j \in \mathbb{Z}_1^J$, the KL divergence is given as
\begin{equation*}
    D_{KL}(q||p) := \sum_{j=1}^{J}q(j)\ln\bigg(\frac{q(j)}{p(j)}\bigg), \,\, q, p \in \Delta_J= \{q \in \mathbb{R}^J\,|\, q \geq 0, \,\sum_{j=1}^{J}q(j) = 1\}.
\end{equation*}
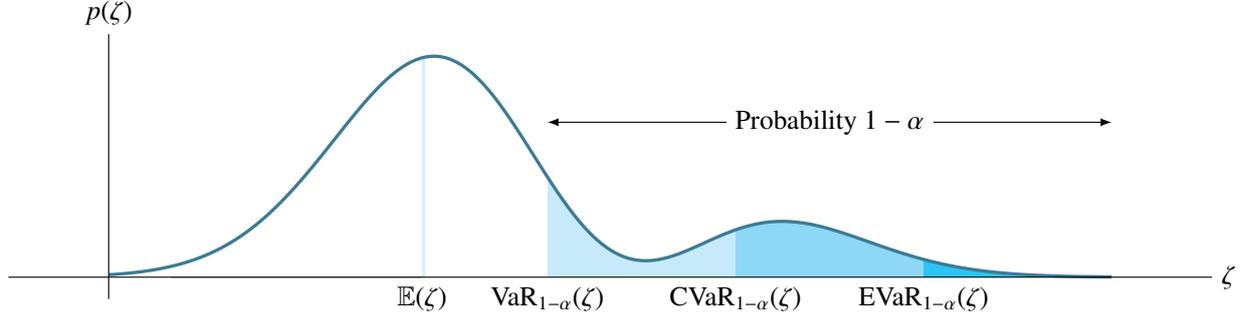
\begin{figure}
\centering{
    \resizebox{1\textwidth}{!}{
\begin{tikzpicture}
\tikzstyle{every node}=[font=\normalsize]
\pgfmathdeclarefunction{gauss}{2}{%
  \pgfmathparse{1000/(#2*sqrt(2*pi))*((x-.5-8)^2+.5)*exp(-((x-#1-6)^2)/(2*#2^2))}%
}

\pgfmathdeclarefunction{gauss2}{3}{%
\pgfmathparse{1000/(#2*sqrt(2*pi))*((#1-.5-8)^2+.5)*exp(-((#1-#1-6)^2)/(2*#2^2))}
}

\begin{axis}[
  no markers, domain=0:16, range=-2:8, samples=200,
  axis lines*=center, xlabel=$\zeta$, ylabel=$p(\zeta)$,
  every axis y label/.style={at=(current axis.above origin),anchor=south},
  every axis x label/.style={at=(current axis.right of origin),anchor=west},
  height=5cm, width=17cm,
  xtick={0}, ytick=\empty,
  enlargelimits=true, clip=false, axis on top,
  grid = major
  ]
   \addplot [fill=cyan!15, draw=none, domain=5:5.05] {gauss(1.5,2)} \closedcycle;
  \addplot [fill=cyan!20, draw=none, domain=7:15] {gauss(1.5,2)} \closedcycle;
    \addplot [fill=cyan!40, draw=none, domain=10:15] {gauss(1.5,2)} \closedcycle;
    \addplot [fill=cyan!65, draw=none, domain=13:15] {gauss(1.5,2)} \closedcycle;
  \addplot [very thick,cyan!50!black] {gauss(1.5,2)};
 
 \pgfmathsetmacro\valueA{gauss2(5,1.5,2)}
 \draw [gray] (axis cs:5,0) -- (axis cs:5,\valueA);
  \pgfmathsetmacro\valueB{gauss2(10,1.5,2)}
  \draw [gray] (axis cs:4.5,0) -- (axis cs:4.5,\valueB);
    \draw [gray] (axis cs:10,0) -- (axis cs:10,\valueB);
 
 
 \draw [gray] (axis cs:1,0)--(axis cs:5,0);
\node[below] at (axis cs:7.0, -0.1)  {$\mathrm{VaR}_{1-\alpha}(\zeta)$}; 
\node[below] at (axis cs:5, -0.1)  {$\mathbb{E}(\zeta)$}; 
\node[below] at (axis cs:10, -0.1)  {$\mathrm{CVaR}_{1-\alpha}(\zeta)$}; 
\node[below] at (axis cs:13, -0.1)  {$\mathrm{EVaR}_{1-\alpha}(\zeta)$};
\draw [yshift=2cm, latex-latex](axis cs:7,0) -- node [fill=white] {Probability~$1-\alpha$} (axis cs:16,0);
\end{axis}
\end{tikzpicture}
}
\caption{Comparison of the mean, VaR, and CVaR for a given confidence $\alpha \in (0,1)$. The axes denote the values of the stochastic variable $\zeta$, i.e., the minimum distance to the safe set as defined in~\eqref{eq:zeta}, and with pdf $p(\zeta)$. The shaded area denotes the $\%(1-\alpha)$ of the area under $p(\zeta)$. If the goal is to minimize $\zeta$, using $\mathbb{E}(\zeta)$ as a performance measure is misleading because tail events with low probability of occurrence are ignored. VaR gives the value of $\zeta$ at the $(1-\alpha)$-tail of the distribution. But, it ignores the values of $\zeta$ with probability below $1-\alpha$.  CVaR is the average of the values of VaR with probability less than $1-\alpha$ (average of the worst-case values of $\zeta$ in the $1-\alpha$ tail of the distribution). Note that $\mathbb{E}(\zeta) \le \mathrm{VaR}_{1-\alpha}(\zeta) \le \mathrm{CVaR}_{1-\alpha}(\zeta) \le \mathrm{EVaR}_{1-\alpha}(\zeta)$. Hence,  $\mathrm{EVaR}_{1-\alpha}(\zeta)$ is a more risk-sensitive measure.}
}
\label{fig:varvscvar}
\end{figure}

\subsection{g-entropic risk measures}

Let $g$ be a convex function with $g(1) = 0$, and $\beta$ be a nonnegative number. The g-entropic risk measure~\cite{ahmadi2012entropic}, $\text{ER}_{g,\beta}$, with divergence level $\beta$ for a random variable $X \in \mathcal{L}_p(\Omega, \mathcal{F}, \mathbb{P})$ is defined as,
\begin{align}\label{eq:gentropic}
    \text{ER}_{g, \beta}(X) := \sup_{Q \in \mathcal{Q}}E_Q(X),
\end{align}
where, $\mathcal{Q} = \{Q \ll P: \int g\big(\frac{dQ}{dP}\big)\, dP \leq \beta\}$.

The definition~\eqref{eq:gentropic} describes the g-entropic risk measures in terms of their dual representation. To obtain the primal form, we can use the generalized Donsker-Vardhan variational formula~\cite{ahmadi2012entropic}, 

\begin{align*}
    \inf_{\mu \in \mathbb{R}} \{\mu + E_P(g^{*}(X - \mu)) \} = \sup_{Q \ll P} \{E_Q(X) - g\big(\frac{dQ}{dP}\big)\, dP \},
\end{align*}
where $g^{*}$ is the conjugate (the Legendre-Fenchel transform) of $g$. Both CVaR and EVaR have been proven to be g-entropic risk measures. Another g-entropic risk measure that we'll consider in this work is the total variation distance~\cite{shapiro2017distributionally}:
\begin{equation*}
    \text{TVD}_\alpha(X) = \sup_{Q \in \mathcal{Q}}E_Q(X) = \alpha\sup_{x\in\Omega}x + (1-\alpha)\text{CVaR}_{1-\alpha}(X)
\end{equation*}
where, for a discrete random variable $X \in \{x_1, x_2,\dotsc, x_J\}$ with pmf given by $p = [p(1), p(2), \dotsc, p(J)]^T$, where $p(j) = \mathbb{P}(X = x_j), j \in \mathbb{Z}_1^J$, the risk envelope is given by,
\begin{equation*}
    \mathcal{Q} := \Big\{q \in \Delta_J: \frac{1}{2}\sum_{j=1}^{J}|q(j)-p(j)|\leq \alpha\Big\}.
\end{equation*}
\section{Problem Statement}~\label{sec:problem_state}
We consider a class of discrete-time dynamical systems given by
\begin{equation}\label{eq:sys}
\begin{aligned}
    \boldsymbol{x}(t+1) &= A\boldsymbol{x}(t) + B\boldsymbol{u}(t) + D\boldsymbol{\delta}(t), \\
    \boldsymbol{y}(t) &= C\boldsymbol{x}(t),
\end{aligned}
\end{equation}
where $\boldsymbol{x}(t) \in \mathbb{R}^{n_x}$, $\boldsymbol{y}(t) \in \mathbb{R}^{n_y}$, and $\boldsymbol{u}(t) \in \mathbb{R}^{n_u}$ are the system state, output, and controls at time $t$, respectively. The system is affected by a stochastic, additive, process noise $\boldsymbol{\delta}(t) \in \mathbb{R}^{n_x}$. In fact, the noise term $\boldsymbol{\delta}$ can represent exogenous disturbances or unmodeled dynamics (see the case study in~\cite{ahmadi2020cvar} for such modeling method applied to bipedal robots). We posit the following assumptions about the availability of measurements and the process noise.
\textcolor{black}{
 \begin{assumption}\label{assumption: measurementavailable}
    A measurement of all states is available at each sample instant and matrix $D$ in Eq. (\ref{eq:sys}) is full-rank.
 \end{assumption}}
\begin{assumption}[Discrete process noise]\label{assumption: process_noise}
\textit{The process \textcolor{black}{noise} $\boldsymbol{\delta}$ is assumed to consist of i.i.d. samples of a discrete distribution given by the probability mass function (pmf), $p_{\delta} = [p_{\delta}(1), p_{\delta}(2), \dotsc, p_{\delta}(J_{\delta})]^T$}. For this distribution, we also define the index set $\mathcal{D} = \mathbb{Z}_1^{J_\delta}$.
\end{assumption}
We also consider $L$ \textit{moving} obstacles with index $l \in\mathbb{Z}_1^L$ that can be approximated by a convex polytope defined by $m_l$ half-spaces in $\mathbb{R}^{n_x}$ 
\begin{equation} \label{eq:obs_def}
    \Bar{\mathcal{O}_l}(t) = \{\boldsymbol{o} \in \mathbb{R}^{n_x} \,|\,\boldsymbol{c}_{i,l}^{T}(\boldsymbol{o} - \boldsymbol{a}_l(t)) \leq \boldsymbol{d}_{i,l}, \, \forall i \in \mathbb{Z}_1^{m_l}\}.
\end{equation}
We allow each polytopic obstacle $\Bar{\mathcal{O}_l}$, $l \in \mathbb{Z}_1^L$, centered at $\boldsymbol{a}_l$ at time $t$ to move randomly w.r.t. the nominal trajectory. The random set defining obstacle $\Bar{\mathcal{O}_l}$, $l \in \mathbb{Z}_1^L$, at time $t$ can be written as a random rotation $R_l$ and random translation $w_l$ of $\Bar{\mathcal{O}_l}$. Hence, we can rewrite the obstacle at time $t$ as a random set, $\mathcal{O}_l$, as
\begin{equation}\label{eq:obs}
\begin{aligned}
    \mathcal{O}_l(t) &= R_l(t)\Bar{\mathcal{O}_l}(t) + \boldsymbol{w}_l(t)\\
    &= \bigg\{\boldsymbol{o'} = R_l(t)(\boldsymbol{o} - \boldsymbol{a}_l(t)) + \boldsymbol{a}_l(t) + \boldsymbol{w}_l(t) \, | \, \boldsymbol{c}_{i,l}^{T}\boldsymbol{o} \leq \boldsymbol{d}_{i,l}, \, \forall i \in \mathbb{Z}_1^{m_l}\bigg\} \\
    &= \bigg\{\boldsymbol{o'} \, | \,\boldsymbol{c}_{i,l}^{T}\Big(R_l(t)^{-1}\big(\boldsymbol{o'}  - \boldsymbol{a}_l(t)  -\boldsymbol{w}_l(t)\big) + \boldsymbol{a}_l(t)\Big)  \leq \boldsymbol{d}_{i,l}, \,\forall i \in \mathbb{Z}_1^{m_l}\bigg \}.
\end{aligned}
\end{equation}
In other words, we allow the $l^{\text{th}}$ obstacle moving along the nominal trajectory $\boldsymbol{a}_l(t)$ to randomly rotate and translate with respect to the nominal trajectory. The random obstacle movement is described by the set $\mathcal{O}_l(t)$ in the above equations.
\begin{figure}
\vspace{-0.3cm}
    \centering
    \includegraphics[width=100mm]{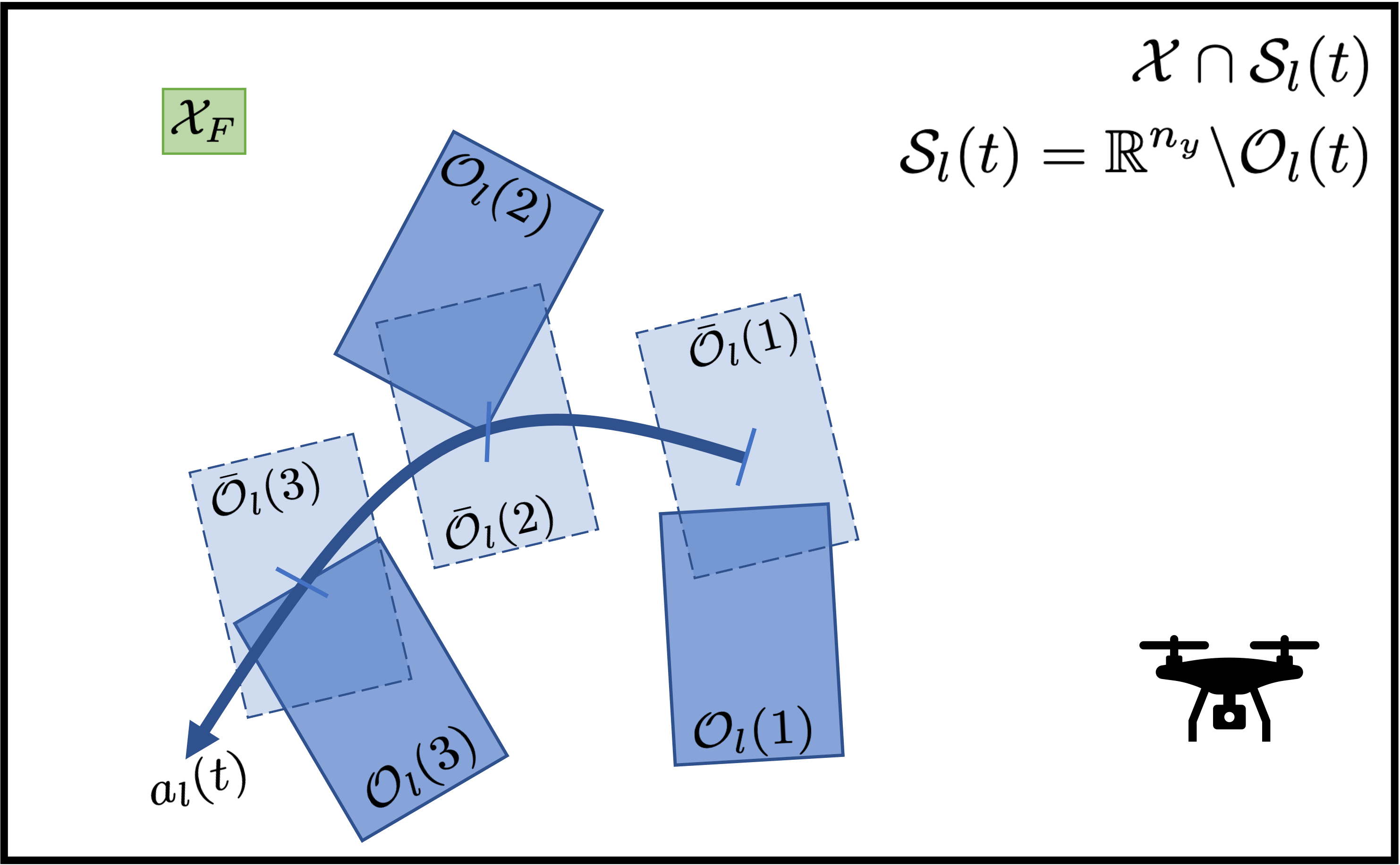}
    \caption{A graphical representation of the motion planning problem. The light blue polytopes $\Bar{\mathcal{O}}(t)$, represent the nominal obstacle set centered at the nominal trajectory $\boldsymbol{a}_l(t)$. We allow for random rotations and translations about this trajectory. This random obstacle set is given by the darker blue polytope $\mathcal{O}(t)$. The safe set, $\mathcal{S}_l(t)$, is the region outside the obstacle set. The goal of the drone in the figure is to plan a path to the terminal set $\mathcal{X}_F$.}
    \label{fig:obs}
\end{figure}

\begin{assumption} [Discrete measurement noise] \label{assumption: measurement_noise}
\textit{The moving obstacles' random rotations and translations relative to a nominal trajectory are sampled from a joint probability distribution whose sample space has cardinality $J_o$, i.e., $\Omega_l = \{(R_l^1, \boldsymbol{w}_l^1), \dotsc, (R_l^J, \boldsymbol{w}_l^{J_o})\}$. A random rotation and translation is picked from this set with pmf given by $p_l = [p_l(1), p_l(2), \dotsc, p_l(J_o)]^T$. For this distribution, we also define the index set $\mathcal{J} = \{1, \dotsc, J_o\}$.}
\end{assumption}
\vspace{0.3cm}

The nominal safe set is defined as the region outside of the polytopic obstacles
\begin{equation}\label{eq:safeset_def}
   \Bar{\mathcal{S}_l}(t) = \mathbb{R}^{n_y} \backslash \Bar{\mathcal{O}_l}(t)
     = \big\{\boldsymbol{o}\,|\, \exists i \in \mathbb{Z}_1^{m_l},~~c_{i,l}^T(\boldsymbol{o} - \boldsymbol{a}_l(t)) \geq d_{i,l}\big\}.
\end{equation}
Similarly, the random safe set is given by $\mathcal{S}_l(t) = \mathbb{R}^{n_y} \backslash \mathcal{O}_l(t)$.
\noindent For obstacle avoidance, we are interested in the minimum  distance to the safe set,
\begin{equation} \label{eq:zeta}
    \zeta(\boldsymbol{y}(t), \mathcal{S}_l(t)) = \text{dist}(\boldsymbol{y}(t), \mathcal{S}_l(t)):= \min_{\boldsymbol{z} \in \mathcal{S}_l(t)} ||\boldsymbol{y}(t) - \boldsymbol{z}||.
\end{equation}

Our goal is to bound the risk of collision with the randomly moving obstacles by evaluating the distance from the probabilistic safe set (which is the complement of the obstacle set) and constraining it to be below a threshold, $\epsilon_l$,
\begin{equation}\label{eq:risksafety}
    \mathrm{\rho}_{1-\alpha}\big[\zeta(\boldsymbol{y}(t), \mathcal{S}_l(t))] \leq \epsilon_l, \quad  \forall l \in \mathbb{Z}_1^L.
\end{equation} 

The obstacle avoidance constraint (\ref{eq:risksafety}) is a risk safety constraint with confidence level $\alpha$ and risk tolerance (also referred to as risk threshold) $\epsilon_l$ for each obstacle~$l \in \mathbb{Z}_1^L$. Note that this implies that we allow the coherent risk of the distance from the safe set to be at most $\epsilon_l$ in $1-\alpha$ worst realizations. Henceforth, we represent $\rho_{1-\alpha}$ as $\rho$ for simplicity. 
 
Let the state constraints take the form $\mathcal{X} := \{x \in \mathbb{R}^{n_x} | F_xx \leq g_x\}, F_x\in \mathbb{R}^{r \times n_x}, g_x \in \mathbb{R}^r$, which can represent physical constraints on a robot. Given that the system is subject to noise $\delta$, we want to satisfy the following state constraints in the risk-averse sense:
\begin{align}\label{eq:state_const}
    \rho(f_{x,n}^Tx(t+k|t) - g_{x,n}) &\leq \epsilon_x, \quad \forall k\in\mathbb{Z}_1^{N}, n\in\mathbb{Z}_1^{r},
\end{align}
where $F_x = \begin{bmatrix} f_{x,1}^T & f_{x,2}^T & \dotsc & f_{x,r}^T\end{bmatrix}^T, g_x = \begin{bmatrix} g_{x,1}^T & g_{x,2}^T & \dotsc & g_{x,r}^T\end{bmatrix}^T$. We write this constraint in shorthand as, $\rho(\boldsymbol{x}_k \not\in \mathcal{X}) \leq \epsilon_x$.

Similarly, we consider control constraints of the form $\mathcal{U} := \{\boldsymbol{u} \in \mathbb{R}^{n_u} | F_u\boldsymbol{u} \leq g_u\}, F_u\in \mathbb{R}^{s \times n_u}, g_u \in \mathbb{R}^s$, for example, representing actuator limitations, and we want to satisfy the following risk constraint
\begin{align}\label{eq:state_const}
    \rho(f_{u,n}^Tu(t+k|t) - g_{u,n}) \leq \epsilon_u, \quad \forall k\in\mathbb{Z}_0^{N-1}, n\in\mathbb{Z}_1^{s},
\end{align}
where $F_u = \begin{bmatrix} f_{u,1}^T & f_{u,2}^T & \dotsc & f_{u,s}^T\end{bmatrix}^T, g_u = \begin{bmatrix} g_{u,1}^T & g_{u,2}^T & \dotsc & g_{u,s}^T\end{bmatrix}^T$. We write this constraint in shorthand as, $\rho(\boldsymbol{u}_k \not\in \mathcal{U}) \leq \epsilon_u$. 

Note that the uncertainty in the control input $\boldsymbol{u}$ arises from the description of the control policy as a function of the disturbances. This disturbance feedback policy will be elaborated upon shortly. If we choose to have hard constraints on the control input, the risk level can be set to a conservative value, $\alpha \rightarrow 1$, for the control constraints. For ease of presentation, we keep the risk level constant across all the constraints. 

Lastly, we also consider terminal constraints of the form $\mathcal{X}_F := \{\boldsymbol{x} \in \mathbb{R}^{n_x} | F_f\boldsymbol{x} \leq g_f\}, F_f\in \mathbb{R}^{v \times n_x}, g_f \in \mathbb{R}^v$ and we want to satisfy,
\begin{align}\label{eq:state_const}
    \rho(f_{f,n}^Tx(t+N|t) - g_{f,n}) \leq \epsilon_f, \quad \forall n\in\mathbb{Z}_1^{v}
\end{align}
where, $F_f = \begin{bmatrix} f_{f,1}^T & f_{f,2}^T & \dotsc & f_{f,v}^T\end{bmatrix}^T, g_f = \begin{bmatrix} g_{f,1}^T & g_{f,2}^T & \dotsc & g_{f,v}^T\end{bmatrix}^T$. We write this constraint in shorthand as, $\rho(\boldsymbol{x}_N \not\in \mathcal{X}_F) \leq \epsilon_f$. 
\begin{rmk}\label{rmk:confidence_adjusted}
The total number of risk constraints are $L+r+s+v$ for $L$ obstacles, $r$ state constraints, $s$  control constraints, and $v$ terminal constraints. With some abuse of notation, we write the risk $\rho_{1-\alpha}$ to mean the adjusted risk level $1-\alpha'$ such that, $$1-\alpha' = \frac{(1-\alpha)}{L+1}$$ where $\alpha$ is the risk confidence level of the entire system and $\alpha'$ is the adjusted risk level for the risk constraints and cost to attain the true confidence $\alpha$.
\end{rmk}
 \begin{assumption}\label{assumption: riskseparable}
 We assume that the measures of risk (used for safety, state, and control constraints and the cost function) are coherent risk measures that can be represented in their dual form as:
 \begin{align*}
    \rho(X) := \sup_{Q \in \mathcal{Q}}E_Q(X),
\end{align*}
where, $\mathcal{Q}$ is a convex, closed set that we represent as $\mathcal{Q} = \big\{ \boldsymbol{g}(q, \alpha) \leq 0,
        \sum_{j=1}^{J}p(j)q(j) = 1,\, q(j)\geq 0, \,\forall j\in\mathcal{J}\big\}$, and $\boldsymbol{g}(q, \alpha)$ is a convex function in $q$. We assume that if $\boldsymbol{g}(q, \alpha)$ is of dimension $>1$, all its elements constitute  a single function applied to all the components of $q$ separately. We also assume that $\alpha \rightarrow 0$ corresponds to the risk-neutral case with $\rho(X) \rightarrow \mathbb{E}(X)$ and $\alpha \rightarrow 1$ corresponds to worst case (robust)  with $\rho(X) \rightarrow \max \delta$.
 \end{assumption}
In this work, we parameterize the control policy as an affine function of the process noise, i.e., we solve for a disturbance feedback policy. \textcolor{black}{As discussed in~\cite{NAVRATIL198863,Oldewurtel2008affineDF}, \textit{open-loop prediction MPC} computes a single set of control actions just as a function of the current state of the system. While this is computationally attractive, this is very conservative because it computes a set of control actions for all possible values of the disturbances that can affect the system instead of accounting for the fact that in the future, the system will have information on all the disturbances that have affected the system so far. Effectively, we are able to choose different responses to the disturbances by parameterizing the control as a \textit{function of the disturbances}. In~\cite{Oldewurtel2008affineDF}, the authors showed that using a policy that is affine in the disturbances is much less conservative and far more  flexible than using open-loop prediction MPC.}

\textcolor{black}{The affine disturbance feedback policy is equivalent to using affine state feedback policies when Assumption~\ref{assumption: measurementavailable} is satisfied. However affine state feedback policies are \textit{nonlinear} in the optimization variables unlike affine disturbance feedback policies that are \textit{linear} in the optimization variables~\cite{GOULART2006523}. This, however, comes at the cost of having more optimization variables in the affine disturbance feedback policy.} Recently, in~\cite{zhang2021SADF}, the authors reduced the number of decision variables \textcolor{black}{from $O(N^2)$ to $O(N)$} for the computation of a typical affine disturbance feedback policy and showed that for a linear, time-invariant system this simplified disturbance feedback policy is still equivalent to computing an affine state feedback policy. We apply this simplified affine disturbance feedback (SADF) as,
\begin{subequations}
\begin{align}\label{eq:controlLaw1}
    u_k &= \sum_{m=0}^{k-1}K_{k-m}\delta_m +\eta_k, \\
    \implies \boldsymbol{u}_{N} &= \boldsymbol{K}_N\boldsymbol{\delta}_{N} + \boldsymbol{\eta}_{N},
\end{align}
\end{subequations}
where, $u_i$ is an affine function of the disturbances, $\delta$, from time $t$ to $t+i$, $K_{k-m}, \eta_k$ are the decision variables in the MPC optimization, and for a $N$ step (horizon) problem,
\begin{align*}
    \boldsymbol{K}_N &= \begin{bmatrix} 0 & \dotsc & \dotsc & 0  & 0\\
                    K_1 & 0 & \dotsc & 0 & 0\\
                    \vdots & \ddots & \ddots & 0 & 0\\
                    K_{N-1} & K_{N-2} & \dotsc & K_1 & 0
    \end{bmatrix} \\ 
    \boldsymbol{u}_{N} &= \begin{bmatrix} u_0 & u_1 & \dotsc & u_{N-1}\end{bmatrix}^T  \\
    \boldsymbol{\delta}_{N} &= \begin{bmatrix} {\delta}_0 & {\delta}_1 & \dotsc & {\delta}_{N-1}\end{bmatrix}^T \\
    \boldsymbol{\eta}_{N} &= \begin{bmatrix} \eta_0 & \eta_1 & \dotsc & \eta_{N-1}\end{bmatrix} ^T.
\end{align*}
Note that \textit{open-loop prediction MPC} is just a subset of affine disturbance feedback policies; this can be observed if we remove the dependence on the feedback gains by setting ${K}_i = 0,\, \, \forall i\in\{1,\dotsc, N-1\}$ and only allowing $\boldsymbol{\eta}_N$ to be the optimization variables. Also note that in closed-loop, we only apply $u_0 = \eta_0$ and that this control input is independent of the disturbances\footnote{\textcolor{black}{This framework can also be adapted for linear, time-varying systems. However, the disturbance feedback policy has to be slightly modified. We cannot use the SADF policy as proposed in~\cite{zhang2021SADF} and have to use a non-simplified version~\cite{nair2022collision}.}}.

\noindent We now present the paper's main problem.
\vspace{0.3cm}
\begin{problem}
\textit{
Consider the discrete-time dynamical system (\ref{eq:sys}) and the randomly moving obstacles $O_l$, $l \in \mathbb{Z}_1^L$, as defined in~\eqref{eq:obs_def} and~\eqref{eq:obs}. Given an initial condition $x_0 \in \mathbb{R}^{n_x}$, a goal set $\mathcal{X}_f \subset \mathbb{R}^{n_x}$, state constraints $\mathcal{X}\subset \mathbb{R}^{n_x}$, control constraints $\mathcal{U}\subset \mathbb{R}^{n_u}$, an immediate convex cost function $r:\mathbb{R}^{n_x} \times \mathbb{R}^{n_u}\to \mathbb{R}_{\ge0}$, a horizon $N \in \mathbb{N}_{\ge 0}$, and risk tolerances $\epsilon_l$, $\epsilon_x, \epsilon_u, \epsilon_F$, for obstacle, state, control and terminal constraints respectively, compute the receding horizon controller $\{u_k \}_{k=0}^{N-1}$ such that $x(t+N) \in \mathcal{X}_f$ while satisfying the risk-sensitive safety constraints~\eqref{eq:risksafety},
\begin{subequations}\label{eq:mpc1}
\begin{align}
\begin{split}
\min_{\boldsymbol{K}_N, \boldsymbol{\eta}_N} \quad &J(x(t), \boldsymbol{u}) := \rho_{1-\alpha}\bigg(\sum_{k=0}^{N-1}(r(\boldsymbol{x}_k, \boldsymbol{u}_k)\bigg) \quad
\end{split}\\
\begin{split}\label{eq:dyn1}
 \textrm{s.t.} \quad &\boldsymbol{x}_{k+1} = A\boldsymbol{x}_k + B\boldsymbol{u}_k + D{\delta}_k,
\end{split}\\
\begin{split}\label{eq:dyn2}
&\boldsymbol{y}_k = C\boldsymbol{x}_k,
\end{split}\\
\begin{split}\label{eq:controlLaw}
&u_k = \sum_{m=0}^{k-1}K_{k-m}\delta_m +\eta_i,
\end{split}\\
\begin{split}\label{eq:stcon}
&\rho_{1-\alpha}(\boldsymbol{x}_{k+1} \not\in \mathcal{X}) \leq \epsilon_x, 
\end{split}\\
\begin{split}\label{eq:contcon}
&\rho_{1-\alpha}(\boldsymbol{u}_k \not\in \mathcal{U}) \leq \epsilon_u, 
\end{split}\\
\begin{split}\label{eq:safetycon}
&\rho_{1-\alpha}\big(\zeta(\boldsymbol{y}_k, \mathcal{S}_l(t+k))\big) \leq \epsilon_l, \forall l \in \mathbb{Z}_1^L, 
\end{split}\\
\begin{split}\label{eq:terminalcon}
\rho_{1-\alpha}(x_N  \not\in \mathcal{X}_F) \leq \epsilon_f,
\end{split}\\
\begin{split}\label{eq:ic}
&\boldsymbol{x}_0 = \boldsymbol{x}(t).
\end{split}
\end{align}
\end{subequations}
}
\end{problem}
\vspace{0.3cm}

Note that although the obstacles $\mathcal{O}_l$ are assumed to be convex polytopes \eqref{eq:obs_def}, the safe set $\mathcal{S}_l(t+k)$ in~\eqref{eq:safeset_def} is nonconvex in $\boldsymbol{y}_k$. Hence, the minimum distance to $\mathcal{S}_l(t+k)$, $\zeta(\boldsymbol{y}_k, \mathcal{S}_l(t+k))$, is also nonconvex in $\boldsymbol{y}_k$. Therefore, the risk-sensitive safety constraint~\eqref{eq:safetycon} is a nonconvex constraint in the decision variable $u$, which renders optimization problem~\eqref{eq:mpc1} nonconvex as well. 

The next section will \textit{reformulate} the state, control, and safety constraints \eqref{eq:stcon}, \eqref{eq:contcon}, \eqref{eq:safetycon} in order to obtain a convex mixed-integer relaxation of \eqref{eq:mpc1}, which yields locally optimal solutions. Nonetheless, every such locally optimal solutions satisfies the constraints of optimization~\eqref{eq:mpc1} including the risk-sensitive safety constraint~\eqref{eq:safetycon}. 
\section{Risk-Constrained Receding Horizon Planning}~\label{sec:mpc_form}
This section breaks down the receding horizon control problem into several parts. First, we modify the state and control constraints by finding efficient approximations that rigorously satisfy the risk constraints. Next, we specifically look at a tractable reformulation of the risk-obstacle avoidance constraint. Note that the risk-averse state and control constraint tightening can be computed offline because it depends only on the risk from the process noise. On the other hand, the risk-averse obstacle avoidance constraint depends on the distance of the system from the obstacle, which is constantly varying and hence needs to be computed online. We then rewrite the non-convex safe set as a set of mixed-integer constraints. We reformulate the terminal constraint by adding discrete states such that we can reach the goal in finite-time. Finally, we provide an efficient, tractable reformulation of the risk cost function. Note that the proofs of the lemmas and propositions are provided in the Appendix.
\subsection{State and Control Constraint Tightening} \label{section: constraint_tighten}

\begin{lem}[Tightened state constraint]\label{lem:stcon_tighten}
Assuming the control policy~\eqref{eq:controlLaw1}, a tightened set of state constraints,
\begin{equation}\label{eq:state_tighten}
\begin{aligned}
     f_{x,n}^T\big(A^{k}x_{0} + \boldsymbol{B}_k\boldsymbol{\eta}_{k}\big) &+ \lVert f_{x,n}^T\big(\boldsymbol{B}_k\boldsymbol{K}_k + \boldsymbol{D}_k\big)\rVert_{1}\rho(|{\delta}|) \leq \epsilon_x + g_{x,n}, \, \forall k\in\mathbb{Z}_1^{N}, n\in\mathbb{Z}_1^{r}    
\end{aligned}
\end{equation}
where, $\boldsymbol{B}_k = \begin{bmatrix}A^{k-1}B & A^{k-2}B & \dotsc & B\end{bmatrix}$, and $\boldsymbol{D}_k = \begin{bmatrix}A^{k-1}D & A^{k-2}D & \dotsc & D\end{bmatrix}$, guarantees that~\eqref{eq:stcon} holds.
\end{lem}
\begin{proof}
   \textcolor{black}{See Appendix A}
\end{proof}
The above tightening is useful because $\rho(|{\delta}|)$ is independent of the disturbance feedback matrices. Hence,  $\rho(|{\delta}|)$ can be computed offline. \textcolor{black}{Without the constraint tightening derived in Lemma~\ref{lem:stcon_tighten}, the state risk constraints have to be evaluated online and this increases the number of MPC constraints by the order of $(J_{\delta})^N$.}

\noindent Similarly, input constraints of the form~\eqref{eq:contcon} are enforced as, 
\begin{align}\label{eq:control_tighten}
    f_{u,n}^T\boldsymbol{\eta_i} + \lVert f_{u,n}^T\boldsymbol{K}_k\rVert_1\rho(|{\delta}|) \leq \epsilon_u + g_{u,n}, & \quad k\in\mathbb{Z}_0^{N-1}, n\in\mathbb{Z}_1^{s},
\end{align}
\noindent and terminal constraints of the form~\eqref{eq:terminalcon} are enforced as, 
\begin{equation}\label{eq:terminal_tighten}
\begin{aligned}
f_{f,n}^T\big(A^{N}x_{0} + \boldsymbol{B}_N\boldsymbol{\eta}_{N}\big) &+ \lVert f_{f,n}^T\big(\boldsymbol{B}_N\boldsymbol{K}_N + \boldsymbol{D}_N\big)\rVert_{1}\rho(|{\delta}|) \leq \epsilon_f + g_{f,n}, \, \forall n\in\mathbb{Z}_1^{v}.
\end{aligned}
\end{equation}

Now we reformulate the risk that arises from the moving obstacle~\eqref{eq:safetycon}. This safety constraint is given by,
\begin{equation}\label{eq:dist_risk}
    {\rho}\big[\zeta(\boldsymbol{y}_k, \mathcal{S}_l(t+k))] = \rho\bigg(\min_{\boldsymbol{z} \in \mathcal{S}_l(t+k)} ||\boldsymbol{y}(t+k|t) - \boldsymbol{z}||\bigg) \leq \epsilon_l, \quad  \forall l \in \mathbb{Z}_1^L
\end{equation}
In~\eqref{eq:dist_risk}, the safe set at time $t+k$, $\mathcal{S}_l(t+k)$, is a random variable that is a function of the discrete measurement noise and the output, $\boldsymbol{y}(t+k|k)$, is a random variable that is a function of the process noise $(\delta_0, \delta_1, \dotsc, \delta_k)$. Hence the distance of the output $\boldsymbol{y}_k$ from the safe set is given by a random variable, $\zeta(\boldsymbol{y}_k, \mathcal{S}_l(t+k))$, that has a joint distribution of the measurement and process noise. This joint distribution has a sample space of cardinality $J = |\mathcal{D}|^k|\mathcal{J}| =(J_{\delta})^kJ_o$ and a pmf given by $\boldsymbol{p}\ = [p(1), p(2), \dotsc, p(J)]^T$. 
\begin{lem}[Safety Constraint Reformulation]\label{lem: safety_constraint}
If Assumptions~\ref{assumption: measurement_noise} and~\ref{assumption: riskseparable} hold, then the L.H.S. of constraint \eqref{eq:safetycon} is equivalent to
\begin{equation}\label{eq: safetycon_reformulate}
\begin{aligned}
    \min_{\boldsymbol{\lambda}_1, \lambda_2, \nu, h_{l,k}} \quad& \lambda_2g^{*}\bigg(\lambda_2^{-1}\big(\boldsymbol{p}(h_{l,k} +\nu)+ \boldsymbol{\lambda}_1\big) \bigg)-\nu&\\
    \text{s.t.}\quad& \boldsymbol{\lambda}_1 \succeq 0, \, \lambda_2 \geq 0,  & \\
     &\lambda_2^{-1}\big(\boldsymbol{p}(h_{l,k}^{j} +\nu)+\boldsymbol{\lambda}_1\big) \in \mathbb{R}^J, &\\
    &\boldsymbol{y}_k +\frac{\boldsymbol{c}_{i,l}}{||\boldsymbol{c}_{i,l}||}h_{l,k}^j \in \mathcal{S}_l^j(t+k), & \forall j \in \mathbb{Z}_1^J,\exists i\in \mathbb{Z}_1^{m_l},
\end{aligned}
\end{equation}
where, $\boldsymbol{\lambda}_1\in\mathbb{R}^J, \lambda_2, \nu, h_{l,k}\in \mathbb{R}$ and $g^*$ is the convex conjugate~\cite{bertsekas2009convex} of the convex function $g$ that describes the risk envelope of a coherent risk measure.
\end{lem}
\begin{proof}
   \textcolor{black}{See Appendix B}
\end{proof}

\begin{rmk}
Note that for simplicity, we are assuming that $g(q)\in\mathbb{R}$. The above proof, however, can easily be extended to a vector-valued function $\boldsymbol{g}$ of dimension $>1$ under Assumption~\ref{assumption: riskseparable}. In this case, $\boldsymbol{g}$ is a function $\Bar{g}$ applied to each component of $q$, i.e.,
\begin{equation*}
    \boldsymbol{g}(q) = \begin{bmatrix} \Bar{g}(q(1)) & \Bar{g}(q(2)) & \dotsc & \Bar{g}(q(J))\end{bmatrix}^T.
\end{equation*}
The equivalent safety constraint reformulation is then given by,
\begin{equation}\label{eq: conjugate_vector}
\begin{aligned}
    \min_{\boldsymbol{\lambda}_1, \boldsymbol{\lambda_2}, \nu, h_{l,k}} \quad&{  \sum_{j\in\mathbb{Z}_1^J}\big\{\lambda^j_2\,\Bar{g}^*\big((\lambda_2^j)^{-1}\big({p(j)}(h_{l,k} +\nu)+ \boldsymbol{\lambda}_1^j\big) \big)\big\}-\nu}\\
    \text{s.t.}\quad& \boldsymbol{\lambda}_1 \succeq 0, \,\boldsymbol{\lambda}_2 \succeq 0,  & \\
    &(\lambda_2^{j})^{-1}\big({p(j)}(h_{l,k}^{j} +\nu)+\boldsymbol{\lambda}^j_1\big) \in \mathbb{R}, &\\
    &\boldsymbol{y}_k +\frac{\boldsymbol{c}_{i,l}}{||\boldsymbol{c}_{i,l}||}h_{l,k}^j \in \mathcal{S}_l^j(t+k),~~\forall j \in \mathbb{Z}_1^J
\end{aligned}
\end{equation}
Similar results are obtained for g-entropic risk measures using the Donsker-Vardhan variational formula, see~\cite{ahmadi2012entropic}. 
\end{rmk}

The above reformulation applies to all coherent risk measures that satisfy Assumption~\ref{assumption: riskseparable}. Next, we present this formulation for a few specific risk measures studied in our examples.

\subsubsection{CVaR}
\noindent For a CVaR constraint, the convex function, $g(q) = \begin{bmatrix}q(1) - \frac{1}{1-\alpha} & \dotsc & q(J) - \frac{1}{1-\alpha}\end{bmatrix}^T$ is separated into $\Bar{g}(q(j)) = q(j) - \frac{1}{1-\alpha}$ $\forall j\in \mathbb{Z}_1^J$. The convex conjugate $\Bar{g}^*(q^*) = \frac{1}{1-\alpha}$ if $q^* = 1$ and $\Bar{g}^*(q^*) = +\infty$ otherwise can be applied to~\eqref{eq: conjugate_vector} and simplified to get a linear program,
\begin{align*}
    \mathrm{CVaR}_{1-\alpha}(\zeta(\boldsymbol{y}_k, \mathcal{S}_l(t+k))) 
    =&\min_{\boldsymbol{\lambda}_1, \boldsymbol{\lambda_2}, \nu}  \sum_{j=1}^J\Bigg\{\lambda^j_2\,\Bar{g}^*\bigg((\lambda_2^j)^{-1}\big({p(j)}(h_{l,k}^{j,*} +\nu)+ \boldsymbol{\lambda}_1^j\big) \bigg)\Bigg\}-\nu \\
    &\quad \text{s.t.} \quad \boldsymbol{\lambda}_1 \succeq 0, \, \boldsymbol{\lambda}_2 \succeq 0 \\
    =&\min_{\boldsymbol{\lambda}_1, \boldsymbol{\lambda_2}, \nu}  \sum_{j=1}^J\frac{\lambda^j_2}{1-\alpha}-\nu \\
    &\quad \text{s.t.} \quad \boldsymbol{\lambda}_1 \succeq 0, \, \boldsymbol{\lambda}_2 \succeq 0, \\
    & \quad \quad \quad \,{p(j)}(h_{l,k}^{j,*} +\nu)+ {\lambda}_1^j = \lambda_2^j, \quad \forall j\in\mathbb{Z}_1^J.
\end{align*}
\subsubsection{EVaR}
\noindent For EVaR, the risk envelope constitutes $g(q) = \sum_{j\in\mathbb{Z}_1^J}p(j)q(j)\ln(q(j)) + \ln(1-\alpha)$ and, 
\begin{equation*}
g^*(q^*) = \sum_{j=1}^Jp(j)\exp\Big({\frac{q^*(j)-1}{p(j)}}\Big) - \ln(1-\alpha)
\end{equation*}
We substitute the convex conjugate into~\eqref{eq:conjugate} to obtain the following exponential cone optimization,
\begin{align*}
   \mathrm{EVaR}_{1-\alpha}(\zeta(\boldsymbol{y}_k, \mathcal{S}_l(t+k))) 
   =&\min_{\boldsymbol{\lambda}_1, \lambda_2, \nu}  \lambda_2g^{*}\bigg(\lambda_2^{-1}\big(\boldsymbol{p}(h_{l,k}^{j,*} +\nu)+ \boldsymbol{\lambda}_1\big) \bigg)-\nu\\
     &\quad \text{s.t.} \quad \boldsymbol{\lambda}_1 \succeq 0, \,\lambda_2 \geq 0  \\
     =&\min_{\boldsymbol{\lambda}_1, \lambda_2, \nu,\boldsymbol{s}}  \lambda_2\sum_{j=1}^J { p(j)\exp\Big({\frac{\lambda_2^{-1}(p(j)(h_{l,k}^{j,*} +\nu)+ {\lambda}_1^j)}{p(j)}}\Big) - \lambda_2\ln(1-\alpha) -\nu}\\
     &\quad \text{s.t.} \quad \boldsymbol{\lambda}_1 \succeq 0 , \,\lambda_2 \geq 0  \\
     =&\min_{\boldsymbol{\lambda}_1, \lambda_2, \nu}  \lambda_2\sum_{j=1}^J  p(j)s(j) - \lambda_2\ln(1-\alpha) -\nu\\
     &\quad \text{s.t.} \quad \boldsymbol{\lambda}_1 \succeq 0, \,\lambda_2 \geq 0,  \\
     & \quad \quad \quad \big(s(j), \,\lambda_2p(j),  \,p(j)(h_{l,k}^{j,*} +\nu)+ {\lambda}_1^j\big) \in K_{exp},\, \forall j\in\mathbb{Z}_1^J.
\end{align*}
\subsubsection{Total variational distance ambiguity sets}
\noindent The risk envelope for TVD is given by, $g(q) = \sum_{j\in\mathbb{Z}_1^J}|q(j)p(j)-p(j)| - 2\alpha$. When the conjugate $g^*(q^*) = \sum_{j\in\mathbb{Z}_1^J}q^*(j) + 2\alpha$ when $|q^*(j)/p(j)| \leq 1 $ is substituted into ~\eqref{eq:conjugate}, one obtains a linear program,

\begin{equation*}
\begin{aligned}
 \mathrm{TVD}_{\alpha}(\zeta(\mathcal{S}_l(t+k))) = 
   =&\min_{\boldsymbol{\lambda}_1, \lambda_2, \nu}  \lambda_2g^{*}\bigg(\lambda_2^{-1}\big(\boldsymbol{p}(h_{l,k}^{j,*} +\nu)+ \boldsymbol{\lambda}_1\big) \bigg)-\nu\\
     &\quad \text{s.t.} \quad \boldsymbol{\lambda}_1 \succeq 0, \,\lambda_2 \geq 0 \\
     =&\min_{\boldsymbol{\lambda}_1, \lambda_2, \nu} \sum_{j=1}^J \big(p(j)(h_{l,k}^{j,*} +\nu)+ {\lambda}_1^j\big) + 2\lambda_2\alpha - \nu\\
     &\quad \text{s.t.} \quad \boldsymbol{\lambda}_1 \succeq 0, \,\lambda_2 \geq 0  \\
     &\qquad \quad |\lambda_2^{-1}\big({p}(j)(h_{l,k}^* +\nu)+ {\lambda}_1^j\big)| \leq p(j),  \quad \,\forall j \in \mathbb{Z}_1^J\\
     =&\min_{\boldsymbol{\lambda}_1, \lambda_2, \nu} \sum_{j=1}^J \big(p(j)(h_{l,k}^{j,*} +\nu)+ {\lambda}_1^j\big) + 2\lambda_2\alpha - \nu \\
     &\quad \text{s.t.} \quad \boldsymbol{\lambda}_1 \succeq 0, \,\lambda_2 \geq 0, \\
       &\qquad \quad -\lambda_2p(j)\leq {p}(j)(h_{l,k}^{j,*} +\nu)+ {\lambda}_1^j \leq \lambda_2p(j),  \quad \,\forall j \in \mathbb{Z}_1^J.
      \end{aligned}
     \end{equation*}
Now that we have seen how our general reformulation of the risk obtained in Lemma~\ref{lem: safety_constraint} can be applied to various examples of coherent risk measures, we are ready to present the MPC optimization that incorporates all the \textcolor{black}{risk constraint reformulation we have obtained from Lemma~\ref{lem: safety_constraint} into the MPC.}
\begin{thm}
\textit{Consider the MPC optimization given by~(\ref{eq:mpc1}) with confidence level $\alpha$ and risk tolerances $\epsilon_l$, $l \in \mathbb{Z}_1^L$. If \textcolor{black}{Assumptions~\ref{assumption: measurementavailable}-\ref{assumption: riskseparable}} hold, then \eqref{eq:mpc1} is equivalent to a minimization over $\mathcal{V} = \{\boldsymbol{K}_N, \boldsymbol{\eta}_N, \boldsymbol{\lambda}_1, {\lambda_2}, \nu, h_{l,k}\}$ given by}
\begin{subequations}\label{eq:mpc2}
\begin{align}
\begin{split} \label{eq:cost2}
\min_{\mathcal{V}} \quad &J(\boldsymbol{x}(t), \boldsymbol{u}) := \rho\Bigg(\sum_{k=0}^{N-1}r(\boldsymbol{x}_k, \boldsymbol{u}_k) \Bigg)
\end{split}\\
\begin{split}
\textrm{s.t.} \quad & \textcolor{black}{\lambda_2 g^*}\bigg(\lambda_2^{-1}\big(\boldsymbol{p}(h_{l,k} +\nu)+ \boldsymbol{\lambda}_1\big) \bigg)-\nu \leq \epsilon_l,
\end{split}\\
\begin{split}
& \boldsymbol{\lambda}_1 \leq 0, \lambda_2 \geq 0,
\end{split}\\
\begin{split}
&\lambda_2^{-1}\big(\boldsymbol{p}(h_{l,k}^{j} +\nu)+\boldsymbol{\lambda}_1\big) \in \mathbb{R}^J,
\end{split}\\
\begin{split} \label{eq:safeset}
& \boldsymbol{y}_k +\frac{\boldsymbol{c}_{i,l}}{||\boldsymbol{c}_{i,l}||}h_{l,k}^j \in \mathcal{S}_l^j(t+k), \quad \forall j\in\mathbb{Z}_1^J
\end{split}\\
\begin{split}
(\ref{eq:dyn1}),(\ref{eq:dyn2}),\textcolor{black}{(\ref{eq:controlLaw})}, (\ref{eq:stcon}), \textcolor{black}{(\ref{eq:contcon}), (\ref{eq:terminalcon})},(\ref{eq:ic}){(\ref{eq:terminalcon})}.
\end{split}
\end{align}
\end{subequations}
\end{thm}
\vspace{0.3cm}
\begin{proof}
We can substitute the result from Lemma~\ref{lem: safety_constraint} in~\eqref{eq:mpc1} to get
\begin{subequations}\label{eq:mpc_minmin}
\begin{align}
\begin{split}
\min_{\boldsymbol{K}_N, \boldsymbol{\eta}_N} \quad &J(\boldsymbol{x}(t), \boldsymbol{u}) := \sum_{k=0}^{N-1}r(\boldsymbol{x}_k, \boldsymbol{u}_k) \quad \end{split}\\
\begin{split}
\textrm{s.t.} \quad & (\ref{eq:dyn1}),(\ref{eq:dyn2}),\textcolor{black}{(\ref{eq:controlLaw})}, (\ref{eq:stcon}), \textcolor{black}{(\ref{eq:contcon}), (\ref{eq:terminalcon})},(\ref{eq:ic}){(\ref{eq:terminalcon})},
\end{split}\\
\begin{split}
\eqref{eq: safetycon_reformulate} \leq \epsilon_l, \quad l \in \mathbb{Z}_1^L.
\end{split}
\end{align}
\end{subequations}

Suppose we have an optimal solution to \eqref{eq:mpc_minmin} given by $(\boldsymbol{K}_N^*, \boldsymbol{\eta}_N^*)$. As \eqref{eq:mpc_minmin} is feasible, its constraints must be satisfied; this implies the inner minimization \eqref{eq: safetycon_reformulate} must also be feasible, with solution $(\boldsymbol{\lambda}_1^*, {\lambda_2}^*, \nu^*, h_{l,k}^*)$. Hence, $(\boldsymbol{K}_N^*, \boldsymbol{\eta}_N^*, \boldsymbol{\lambda}_1^*, {\lambda_2}^*, \nu^*, h_{l,k}^*)$ must also be a feasible solution to \eqref{eq:mpc2} and yield the same objective value. Conversely, denote the optimal solution to \eqref{eq:mpc2} as $(\boldsymbol{K}_N^*, \boldsymbol{\eta}_N^*, \boldsymbol{\lambda}_1^*, {\lambda_2}^*, \nu^*, h_{l,k}^*)$. The pair $(\boldsymbol{K}_N^*, \boldsymbol{\eta}_N^*)$ must be feasible for \eqref{eq:mpc_minmin} and gives the same objective value. Hence, the above optimization \eqref{eq:mpc_minmin} is equivalent to the one-layer optimization~\eqref{eq:mpc2}.
\end{proof}

\textcolor{black}{We have now included the results from Lemma~\ref{lem: safety_constraint} into the MPC formulation to get an equivalent formulation~\eqref{eq:mpc2} of the original MPC problem given by~\eqref{eq:mpc1}. We can incorporate the results from Lemma~\ref{lem:stcon_tighten} by replacing constraints~\eqref{eq:stcon},~\eqref{eq:contcon},~\eqref{eq:terminalcon} with the tightened constraints~\eqref{eq:state_tighten},~\eqref{eq:control_tighten},~\eqref{eq:terminal_tighten} respectively. However, it remains to express the cost~\eqref{eq:cost2} and obstacle avoidance constraint~\eqref{eq:safeset} just in terms of the optimization variables $\boldsymbol{K}_N, \boldsymbol{\eta}_N$ instead of the dependence on $\boldsymbol{x}_k,\boldsymbol{u}_k$. We will reformulate the obstacle avoidance constraint in terms of the optimization variables and provide a mixed-integer reformulation of the nonconvex safe set in the next subsection.}
\subsection{Mixed-Integer Reformulation of the MPC optimization}

The nonconvex safe set can be described as a set of disjunctive inequalities, which are incorporated in our optimization by introducing a set of binary variables and using the Big-M reformulation~\cite{vecchietti_modeling_2003}.
The safe set~\eqref{eq:safeset_def} is defined as the region outside the obstacle $l$. Given that an obstacle can rotate and translate by $R_l(t+k)$ and $\boldsymbol{w}_l(t+k)$ from its nominal trajectory, we can write the safe set at $t+k$ as the region outside $\mathcal{O}_l(t+k)$ described in \eqref{eq:obs}. It can equivalently be expressed as a result of the rotation and translation of the nominal safe set itself
\begin{equation}
    \mathcal{S}_l(t+k)  = \mathbb{R}^{n_y} \backslash \mathcal{O}_l(t+k) 
     = R_l(t+k)\Bar{\mathcal{S}_l}(t+k) + \boldsymbol{w}_l(t+k).
\end{equation}

In (\ref{eq:safeset}), $\mathcal{S}_l(t+k)$ is a nonconvex set. For some obstacle~$l \in \mathbb{Z}_1^L$,~\eqref{eq:safeset} can be rewritten as
\begin{equation*}
    R_l(t+k)^{-1}\bigg(\boldsymbol{y}_k +\frac{\boldsymbol{c}_{i,l}}{||\boldsymbol{c}_{i,l}||}h_{l,k} - \boldsymbol{w}_l(t+k)\bigg) \in \Bar{\mathcal{S}_l}(t+k).
\end{equation*}
We know that $\boldsymbol{y}_k = C\boldsymbol{x}_k$ such that $\boldsymbol{x}_k = A^{k}x_{0} + \boldsymbol{B}_k(\boldsymbol{\eta}_{k} + \boldsymbol{K}_k\boldsymbol{\delta}_{k} + \boldsymbol{D}_k\boldsymbol{\delta}_{k})$, where $\boldsymbol{\delta}_k = \begin{bmatrix}{\delta}_1 & {\delta}_2 & \dotsc {\delta}_k \end{bmatrix}^T$ is the process noise, i.e., 
\begin{align*}
    \small{R_l(t+k)^{-1}\big( A^{k}x_{0} + \boldsymbol{B}_k(\boldsymbol{\eta}_{k} + \boldsymbol{K}_k\boldsymbol{\delta}_{k}) + \boldsymbol{D}_k{\delta}_{k}  + \frac{\boldsymbol{c}_{i,l}}{||\boldsymbol{c}_{i,l}||}h_{l,k} - \boldsymbol{w}_l(t+k)\big) \in \Bar{\mathcal{S}_l}(t+k).}
\end{align*}
In the above equation, 
the safe set at time $t+k$, $\mathcal{S}_l(t+k)$, is a random variable that is a function of the discrete measurement noise and the output, $\boldsymbol{y}(t+k|k)$, that in turn is a random variable that is a function of the process noise $(\delta_0, \delta_1, \dotsc, \delta_k)$. Hence, distance of the output from the safe set, $h_{l,k}$, is a random variable that has a joint distribution of the measurement and process noise. This joint distribution has a sample space of cardinality $J = |\mathcal{D}|^k|\mathcal{J}| =(J_{\delta})^kJ_o$ and a pmf given by $\boldsymbol{p}\ = [p(1), p(2), \dotsc, p(J)]^T$. 
\begin{align*}
    \small{R_l^j(t+k)^{-1}\big( A^{k}x_{0} + \boldsymbol{B}_k\boldsymbol{\eta}_{k} + \big(\boldsymbol{B}_k\boldsymbol{K}_k + \boldsymbol{D}_k\big)\boldsymbol{\delta}^j_{k} + \frac{\boldsymbol{c}_{i,l}}{||\boldsymbol{c}_{i,l}||}h_{l,k}^j - \boldsymbol{w}_l^j(t+k)\big) \in \Bar{\mathcal{S}_l}^j(t+k),}
\end{align*}
where $\{R_l^j(t+k), \,{w}_l^j(t+k), \,\boldsymbol{\delta}^j_{k}\}$ such that $j \in \mathbb{Z}_1^J$ is a realization of the measurement and process noise from its joint distribution. Given that the obstacles are convex polygons of the form~\eqref{eq:obs_def}, we write the safe region as the union of regions outside of the halfspaces that define an obstacle as follows
\begin{multline}\label{eq:disjunction}
   \bigvee_{i=1}^{m_l} \boldsymbol{c}_{i,l}^T \bigg[ R_l^j(t+k)^{-1}\big( A^{k}x_{0} + \boldsymbol{B}_k\boldsymbol{\eta}_{k} + \big(\boldsymbol{B}_k\boldsymbol{K}_k + \boldsymbol{D}_k\big)\boldsymbol{\delta}^j_{k} + \frac{\boldsymbol{c}_{i,l}}{||\boldsymbol{c}_{i,l}||}h_{l,k}^j - \boldsymbol{w}_l^j(t+k) - \boldsymbol{a}_l(t+k)\big) + \boldsymbol{a}_l(t+k) \bigg] \geq d_{i,l}.
\end{multline}
Because the above disjunctive inequalities, however, are hard to enforce, we reformulate the constraint using a Big-M reformulation. The reformulation converts the disjunctive inequalities into a set of constraints described using binary variables, $\gamma_i^j \in \{0,1\}$ and a large positive constant $M$. The value of $M$ depends on the bounds on $h_{l,k}^j$ (determined from the size of obstacle $l$) and $\boldsymbol{y}_k$ (dependent on the state and control inputs). It can be computed using linear programming. The Big-M reformulation of \eqref{eq:disjunction} is as follows

\begin{subequations} \label{eq:bigM}
  \begin{align}
    \begin{split}\label{eq:bigM1} 
     \boldsymbol{c}_{i,l}^T \Bigg[ R_l^j(t+k)^{-1}\bigg(A^{k}x_{0} + \boldsymbol{B}_k\boldsymbol{\eta}_{k} + \big(\boldsymbol{B}_k\boldsymbol{K}_k + \boldsymbol{D}_k\big)\boldsymbol{\delta}^j_{k} + \frac{\boldsymbol{c}_{i,l}}{||\boldsymbol{c}_{i,l}||}h_{l,k}^j - \boldsymbol{w}_l^j(t+k) - \boldsymbol{a}_l(t+k)\bigg)  + \boldsymbol{a}_l(t+k) \Bigg]  \geq d_{i,l} - M\gamma_i^j,
    \end{split}\\
    \begin{split}
        \sum_{i = 1}^{m_l} \gamma_i^j \leq m_l - 1 , \hspace{50pt}\forall i \in \mathbb{Z}_1^{m_l},\, j \in \mathbb{Z}_1^{J}.
    \end{split}
   \end{align}
\end{subequations}
Inequalities~\eqref{eq:bigM} provide output constraints that satisfy the risk-sensitive obstacle avoidance constraint by taking into account measurement noise and process noise. However, the cardinality of the joint distribution that describes the distance from the obstacle, $h_{l,k}$, increases exponentially with the horizon, $k$. With an exponentially increasing number of mixed-integer variables, the optimization soon becomes intractable. 

To account for this, we introduce a new random variable, $\delta_{max,k} $, whose cumulative distribution function is defined as follows
\begin{subequations} 
\begin{align}
    \mathbb{P}(\delta_{max,k} \leq x) :=& \mathbb{P}(|\delta_0| \leq x \, ,\, |\delta_2| \leq x \, , \, \dotsc \, ,\, |\delta_{k-1}| \leq x) \\
    =& \mathbb{P}(|\delta_0| \leq x)\mathbb{P}(|\delta_2| \leq x)\dotsc\mathbb{P}(|\delta_{k-1}| \leq x) \\
    =& \mathbb{P}(|\delta_1| \leq x)^k.
\end{align}
\end{subequations}
We can find a conservative (inner) approximation of~\eqref{eq:bigM1} using, ${\delta}_{max,k}$, as follows,
\begin{align*}
    \boldsymbol{c}_{i,l}^T R_l(t+k)^{-1}\big(\boldsymbol{B}_k\boldsymbol{K}_k + \boldsymbol{D}_k\big)\boldsymbol{\delta}_{k} \geq& -\lVert\boldsymbol{c}_{i,l}^T R_l(t+k)^{-1}\big(\boldsymbol{B}_k\boldsymbol{K}_k + \boldsymbol{D}_k\big)\rVert_1\lVert\boldsymbol{\delta}_{k}\rVert_{\infty} \\
    =& -\lVert\boldsymbol{c}_{i,l}^T R_l(t+k)^{-1}\big(\boldsymbol{B}_k\boldsymbol{K}_k + \boldsymbol{D}_k\big)\rVert_1\delta_{max,k}.
\end{align*}
Hence, we can rewrite~\eqref{eq:bigM1} as,
\begin{equation}\label{eq: bigM_max_approx}
\begin{split}
    -\big\lVert\boldsymbol{c}_{i,l}^T& R_l^j(t+k)^{-1}(\boldsymbol{B}_k\boldsymbol{K}_k + \boldsymbol{D}_k)\big\rVert_1\delta_{max,k}^j + 
    \boldsymbol{c}_{i,l}^T \bigg[ R_l^j(t+k)^{-1}\Big(A^{k}x_{0} +\boldsymbol{B}_k\boldsymbol{\eta}_{k} +  \frac{\boldsymbol{c}_{i,l}}{||\boldsymbol{c}_{i,l}||}h_{l,k}^j - \boldsymbol{w}_l^j(t+k)- \boldsymbol{a}_l\Big) + \boldsymbol{a}_l\bigg]  \geq d_{i,l} - M\gamma_i^j
\end{split}
\end{equation}
 Notice that in the above inequality, we have, with some abuse of notation, reduced the cardinality of the joint distribution that describes, $h_{l,k}$, from $J = |\mathcal{D}|^k|\mathcal{J}| =(J_{\delta})^kJ_o$ to $J = |\mathcal{D}||\mathcal{J}| =J_{\delta}J_o$. This means that the number of constraints no longer increase exponentially with horizon length. Notice that the approximation is not conservative at the beginning of the horizon, i.e., when $k=1$, the distribution described by $\delta_{max,k}$ is the same as $\delta_1$. 
\subsection{Terminal constraints}
In order to steer the system to the target region in finite time, we follow the suggestion of \cite{richards2003robustFeasibility} and define a new discrete state $\psi \in \{ 0, 1\}$, such that $\psi = 0$ implies that the task has been completed at an earlier step or at the current step and $\psi = 1$ means that the task has not yet been completed at the current time. The update equation of $\psi$ is then given by
\begin{equation}\label{eq:taskCompletion}
    \psi_{k+1} = \psi_{k} - \mu_{k},
\end{equation}
\noindent where ${\mu}_{k} \in \{ 0, 1\}$ is a discrete input. 

Our goal to drive the system to a terminal set $\mathcal{X}_F$ is given by the tightened state constraint~\eqref{eq:terminal_tighten}. Additionally, we incorporate the following constraints
\begin{equation}\label{eq:invariance_states}
\begin{aligned}
    f_{f,n}^T\big(A^{k}x_{0} + \boldsymbol{B}_k\boldsymbol{\eta}_{k}\big) &+ \lVert f_{f,n}^T\big(\boldsymbol{B}_k\boldsymbol{K}_k + \boldsymbol{D}_k\big)\rVert_{1}\rho(|{\delta}|) \leq \epsilon_f + g_{f,n} + \mathbbm{1}{M}(1 - \mu_{k}),
\end{aligned}
\end{equation}
$\forall k\in\mathbb{Z}_1^{N}, n\in\mathbb{Z}_1^{v}$, where $\mathbbm{1}\in\mathbb{R}^{n_x}$ is a vector of $1$'s. Here, ${\mu}_{k} = 1$ if the task of reaching the goal, $\mathcal{X}_F = \{\boldsymbol{x}\in\mathbb{R}^{n_x} | F_f\boldsymbol{x} \leq g_f\}$ is completed at time step $t+k+1$. Equation \eqref{eq:taskCompletion} implies that $\psi$ jumps from $1 \rightarrow 0$, signaling completion of the task. After the task completion, all other MPC problem constraints can be relaxed by adding the term $M(1 - \psi_{k})$ to them, i.e., any constraints of the form $C_1\boldsymbol{s}_k + C_2\gamma_i + C_3 \geq 0$ are modified to $C_1\boldsymbol{s}_k + C_2\gamma_i  + C_3 + \mathbbm{1}M(1 - \psi_{k})\geq 0, \, \forall i, k$ where $\boldsymbol{s}_k = [\boldsymbol{K_k,\eta_k,\lambda_1, \lambda_2},\nu, h_{l,k}] $. We also add the following terminal constraint at time $t + N$ to ensure that the task is completed
\begin{equation} \label{eq:terminalConst} 
    \psi_{N} = 0.
\end{equation}

Note that the discrete state $\psi$ need not be a binary variable as long as we enforce the constraint,
\begin{equation}\label{eq:discreteStateConst}
    0 \leq \psi_{k} \leq 1, \quad k=1,2,\ldots,N.
\end{equation}

The MPC objective function is then modified as
\begin{equation}\label{eq:MPC_cost1}
\begin{aligned}
    \min_{\mathcal{V}} \quad J(\mathcal{V}) :=& \,\rho\bigg(\sum_{k=0}^{N-1}\big(r(\boldsymbol{u}_k) + \psi_k\big)\bigg) \\
    =& \,\rho\bigg(\sum_{k=0}^{N-1}r\big(\sum_{m=0}^{k-1}K_{k-m}\delta_m +\eta_k\big)\bigg) + \sum_{k=0}^{N-1}\psi_k \\
\end{aligned}
\end{equation}
where $\mathcal{V} = \{\boldsymbol{K_N,\eta_N,\lambda_1, \lambda_2},\nu, h_{l,k}\}$ and $r(\boldsymbol{u}_k)$ is a convex function of the control input with $r(0) = 0$.

\subsection{MPC Objective}
 For the MPC cost~\eqref{eq:MPC_cost1}, consider the case, $r(\boldsymbol{u}) = \lVert R\boldsymbol{u}\rVert_1$, where $R \in \mathbb{R}^{n_u}$, 
\begin{equation}\label{eq:obj}
\begin{aligned}
    J(t) &= \rho\Big(\sum_{k=0}^{N-1}\lVert R\boldsymbol{u}_k\rVert_1\Big) + \sum_{k=0}^{N-1}\psi_k, \\ &= \rho\bigg(\sum_{k=0}^{N-1}\big\lVert\sum_{m=0}^{k-1}RK_{k-m}\delta_m +R\eta_k\big\rVert_1\bigg) + \sum_{k=0}^{N-1}\psi_k.
\end{aligned}
\end{equation}
This subsection introduces two methods to compute the control effort risk given by $\rho\big(\sum_{k=0}^{N-1} \,r(\boldsymbol{u_k})\big)$. The first method provides an exact value of the risk and the second method provides an approximation. The first method loosely follows the steps taken to calculate the moving obstacle risk (see Lemma~\ref{lem: safety_constraint}) and will be more computationally expensive because the control effort $r(\boldsymbol{u}_k)$ is a joint distribution of $(\delta_0, \dotsc, \delta_{N-1})$ that grows with the horizon length $N$. The second method will utilize the constraint tightening tools used in Lemma~\ref{lem:stcon_tighten} to approximate the value of the control effort risk. This approximation will be more computationally efficient. The examples in Section~\ref{sec:results} contrast the two methods.
\subsubsection{Exact computation of control effort risk}
\noindent We define the control effort as a random variable, $Z:= \sum_{k=0}^{N-1}\big\lVert\sum_{m=0}^{k-1}K_{k-m}\delta_m +\eta_k\big\rVert_1$. The sample space of $Z$ consists of the joint probability distribution of $(\delta_0, \dotsc, \delta_{N-1})$, which has cardinality $|\mathcal{D}|^N$. All the realizations of $Z$ can be vectorized as $\boldsymbol{z} = \begin{bmatrix} z(1), z(2),\dotsc, z(|\mathcal{D}|^N)\end{bmatrix}$. Note that $z(j) = \sum_{k=0}^{N-1}\big\lVert\sum_{m=0}^{k-1}K_{k-m}\delta_m^j +\eta_k\big\rVert_1,~ \forall j\in \mathbb{Z}_1^{|\mathcal{D}|^N}$, where $\delta_m^j$ is a realization of $\delta_m$ from the joint pmf. If the pmf is denoted by $\boldsymbol{p}_{\Delta}\in\mathbb{R}^{|\mathcal{D}|^N}$, then,
\begin{equation}\label{eq:exact_cost}
\begin{aligned}
   \rho\bigg(\sum_{k=0}^{N-1}\big\lVert \sum_{m=0}^{k-1}RK_{k-m}\delta_m +R\eta_k\big\rVert_1\bigg) &= \left \lbrace\begin{matrix} \max_{q(1), \dotsc, q(|\mathcal{D}|^N)} & \mathbb{E}_Q\big[Z\big] & \\
        \text{s.t.} & {g}(q) \leq 0 & \\
        &-q(j)\leq 0 & \forall j\in\mathbb{Z}_1^{|\mathcal{D}|^N} \\
        &\sum_{j=1}^{J}p_{\Delta}(j)q(j) = 1 &  \end{matrix} \right. \\
        &= \left \lbrace\begin{matrix}  \min_{\boldsymbol{\xi}_1, \xi_2, \vartheta} \quad& \xi_2g^{*}\bigg(\xi_2^{-1}\big(\boldsymbol{p}_{\Delta}(\boldsymbol{z} +\vartheta)- \boldsymbol{\xi}_1\big) \bigg)-\vartheta&\\
    \text{s.t.}\quad& \boldsymbol{\xi}_1 \leq 0  & \\
    & \xi_2 \geq 0  & ,\end{matrix} \right.\\
    \end{aligned}
\end{equation}
where, $\boldsymbol{\xi}_1 \in\mathbb{R}^{|\mathcal{D}|^N}, \, \boldsymbol{\xi}_2, \vartheta \in \mathbb{R}$ are the dual variables, see Lemma~\ref{lem: safety_constraint} for details on how we find the dual function and obtain the conjugate in the above minimization. Note that the number of constraints grow exponentially with the horizon length.

\subsubsection{Approximation of control effort risk}
\noindent We can alternatively approximate the cost as, 
\begin{equation}\label{eq:upper_bound_cost}
\begin{aligned}
    &\rho\bigg(\sum_{k=0}^{N-1}\big\lVert \sum_{m=0}^{k-1}RK_{k-m}\delta_m +R\eta_k\big\rVert_1\bigg) \\&\leq \sum_{k=0}^{N-1}\Big(\rho\big(\lVert \sum_{m=0}^{k-1}(RK_{k-m}\delta_m)\rVert_1\big) + \lVert R\eta_k\rVert_1\Big) & {\text{(Subadditivity)} }\\
     &\leq \sum_{k=0}^{N-1}\Big(\big\lVert \sum_{m=0}^{k-1} RK_{k-m}\big\rVert_1\rho\big( |\delta|\big) + \lVert R\eta_k\rVert_1\Big) & {\text{(i.i.d disturbances)}}, \\
\end{aligned}
\end{equation}
where we obtained the first inequality by using the subadditivity of norms and then the translational invariance property of coherent risk measures. The second inequality results from observing that all disturbances are i.i.d and can be replaced by $\delta$. We use the homogeneity of norms and coherent risk measures to obtain the final result (similar to Lemma~\ref{lem:stcon_tighten}).

The above cost approximation eliminates the additional $|\mathcal{D}|^N$ constraints that result from~\eqref{eq:exact_cost}. This approximation deprioritizes task completion; i.e., when we substitute this upper bound into~\eqref{eq:obj}, the term $\sum_{k=0}^{N-1}\psi_k$ has less weight compared to when we use the exact cost~\eqref{eq:exact_cost}. Another approximation of the true cost would be using the random variable $\delta_{max, N}$ as seen in~\eqref{eq: bigM_max_approx}. In using an approximation of $(\delta_0, \dotsc, \delta_{N-1})$ via $\delta_{max, N}$, we reduce the number of constraints from $|\mathcal{D}|^N$ to $|\mathcal{D}|$.

\subsection{Properties of MPC}
We now combine all the parts of the MPC into one optimization.

\begin{equation}\label{eq:mpc_exact}
\begin{aligned}
\min_{\mathcal{V}} \quad& \sum_{k=0}^{N-1}\Big(\big\lVert \sum_{m=0}^{k-1} RK_{k-m}\big\rVert_1\rho\big( |\delta|\big) + \lVert R\eta_k\rVert_1\Big) &\\
    &\sum_{j=1}^{J}\Bigg\{\lambda^j_2\,\Bar{g}^*\bigg((\lambda_2^j)^{-1}\big({p(j)}(h_{l,k}^j +\nu)+ \boldsymbol{\lambda}_1^j\big) \bigg)\Bigg\}-\nu \leq \epsilon_l + M_k&\\
    &\text{L.H.S}~\eqref{eq: bigM_max_approx}\geq d_{i,l} - M\gamma_i^j- M_k\\
    & F_x\big(A^{k}x_{0} + \boldsymbol{B}_k\boldsymbol{\eta}_{k}\big) + \lVert F_x\big(\boldsymbol{B}_k\boldsymbol{K}_k + \boldsymbol{D}_k\big)\rVert_{1}\rho(|{\delta}|)  \leq \epsilon_x + g_x+ M_k\\
    & F_u\boldsymbol{\eta_k} + \lVert F_u\boldsymbol{K}_k\rVert_1\rho(|\boldsymbol{\delta}|)\big) \leq \epsilon_u + g_u+ M_k\\
    &\eqref{eq:taskCompletion},~\eqref{eq:invariance_states},~\eqref{eq:terminalConst},~\eqref{eq:discreteStateConst}.
\end{aligned}
\end{equation}
where $M_k = M(1 - \psi_{k})$ and $\mathcal{V} = \{\boldsymbol{K}_N,\boldsymbol{\eta}_N,\boldsymbol{\lambda}_1, \boldsymbol{\lambda}_2,\nu, h_{l,k}\}$. The constraints must hold $\forall k\in \mathbb{Z}_1^{N-1},\, l\in \mathbb{Z}_1^L,\, j \in \mathbb{Z}_1^J,$ and $ i \in \mathbb{Z}_1^{m_l}$.
The solution to the  deterministic MPC problem~\eqref{eq:mpc_exact} is also a solution to~\eqref{eq:mpc1}.
The convex, mixed-integer relaxation of a nonconvex optimization problem in~\eqref{eq:mpc_exact} results in locally optimal solutions. 

\begin{proposition}[Risk-sensitive recursive feasibility]\label{proposition: rec_feasibility}
 \textcolor{black}{Assume that} optimization~\eqref{eq:mpc_exact} is feasible at time $t$, \textcolor{black}{then ~\eqref{eq:mpc_exact}} is feasible at time $t+1$ with confidence $\alpha$. 
\end{proposition}
\begin{proof}
   \textcolor{black}{See Appendix C}
\end{proof}

\begin{rmk}
Proposition~\ref{proposition: rec_feasibility} provides a loose bound on the infeasibility of the MPC. In the case of CVaR, TVD, and EVaR, we know that the risk measures are upper bounds for VaR (see Figure 1) and can hence provide tighter bounds on the likelihood of infeasibility. Recall from Remark~\ref{rmk:confidence_adjusted} that the confidence level is adjusted to account for multiple risk constraints. Hence, we can quantify the bounds on probability of MPC infeasibility in terms of the confidence $\alpha$,
\begin{equation*}
    \begin{aligned}
     \mathbb{P}\{\text{MPC infeasible \textcolor{black}{at }}\textcolor{black}{t+1 | t}\}
      &\leq \mathbb{P}\{\delta_0 > \rho(|\delta|) \, \cup\, h_{l,0} > \rho(h_{l,0})\} \\
      &\leq  \mathbb{P}\{\delta_0 \geq \rho(|\delta|)\} + \sum_{l=1}^{L}\mathbb{P}\{h_{l,0} \geq \rho(h_{l,0})\} \\
      &\leq \frac{1-\alpha}{L+1} + L\frac{1-\alpha}{L+1} \qquad {\text{(VaR probability bound)}}\\
      &\leq 1 - \alpha.
    \end{aligned}
\end{equation*}
\end{rmk}
MPC is often used to plan local trajectories given a reference trajectory or a set of waypoints from a higher-level global planner like A* or RRT ~\cite{Lopez2017, hakobyan2019cvar}. Let  $\{\boldsymbol{w}_1, \boldsymbol{w}_2, \dotsc, \boldsymbol{w}_K\}$ be a given a sequence of waypoints. We call a waypoint $\boldsymbol{w}_{j+1}$ \textit{$N$-step reachable} from $\boldsymbol{w}_j$, if there exists a feasible solution to \eqref{eq:mpc_exact} with $\boldsymbol{x}_0 = \boldsymbol{w}_j$ and $\boldsymbol{x}_{K} = \boldsymbol{w}_{j+1}$.\footnote{We assume that we obtain these waypoints from a higher-level planner like A* or RRT. Analyzing the $N$-step reachability of the waypoints is out of the scope of this paper and we consider it an avenue of future work.}

\begin{algorithm}[t!]
\caption{Follow waypoints}\label{alg:waypoint_alg}
\begin{algorithmic}
    \STATE Number of waypoints visited, $W = 0$
    \WHILE{$W < K$}
        \STATE Initialize $(\boldsymbol{x}_0, \psi_0) = (\boldsymbol{w}_{W}, 1)$
        \STATE Set desired goal $\boldsymbol{x}_{des} = \boldsymbol{w}_{W+1}$  
        \WHILE{$\psi_0 \neq 0$}
            \STATE Solve \eqref{eq:mpc_exact} to obtain policy ${\{(0, \eta_0^*),\dotsc , ({K}_{N-1}^*,\eta_{N-1}^*)\}}$ 
            \STATE Update $\boldsymbol{x}_0 = A\boldsymbol{x}_0 + B \eta_0^* + D\delta_0$ 
            \STATE Update $\psi_0 = \psi_0 - \mu_0$ 
            \IF{$\boldsymbol{x}_0 =\boldsymbol{x}_{des}$}
                \STATE $W = W + 1$
            \ENDIF
        \ENDWHILE
    \ENDWHILE
    \end{algorithmic}
\end{algorithm}
\begin{proposition}[Finite-time task completion]\label{proposition:task_completion}
 \textit{Assuming that the waypoint $\boldsymbol{w}_{j+1}$ is $N$-step reachable from $\boldsymbol{w}_j, \, \forall j \in \mathbb{Z}_0^{K-1}$, Algorithm \ref{alg:waypoint_alg} gives a sequence of control inputs to move from $\boldsymbol{w}_0$ to $\boldsymbol{w}_K$ in finite time with confidence \textcolor{black}{$\alpha^{J_0^{K-1}}$, where ${J_0^{K-1}} = \sum_{j=0}^{K-1} \lceil J^{*}_{w_j} \rceil$ and $J^{*}_{w_j}$ refers to the cost of the MPC optimization~\eqref{eq:mpc_exact} to reach waypoint $\boldsymbol{w}_{j+1}$ at the time-step after waypoint $\boldsymbol{w}_{j}$ has been reached.}}
\end{proposition}
\begin{proof}
   \textcolor{black}{See Appendix D}
\end{proof}

\section{Numerical Results}\label{sec:results}
To illustrate the effectiveness of this method, we present numerical experiments that were run on MATLAB using the YALMIP toolbox~\cite{yalmip} with a Gurobi solver~\cite{gurobi} (for CVaR and TVD) and a Mosek solver~\cite{mosek} (for EVaR). 
\subsection{Simple 2D system}
We first look at the two-dimensional discrete system $ x_{k+1} = Ax_k + Bu_k + D\delta_k$ that is similar to the example we considered in~\cite{dixit2020risksensitive}, but with process noise.
\begin{equation*}{\small
   A = \begin{bmatrix} 1.0475 & -0.0463 \\ 0.0463 & 0.9690\end{bmatrix}, \, B = \begin{bmatrix} 0.028 \\ -0.0195\end{bmatrix}, \, D = \begin{bmatrix} 0.028 \\ -0.0195\end{bmatrix}.}
\end{equation*}
The process noise can take values, $\delta_k \in \begin{bmatrix}-0.2 & -0.1 & 0 & 0.1 & 0.2\end{bmatrix}. \\$
The control constraints are $$-100 \leq u_k \leq 100.$$  One randomly moving obstacle interferes with the MPC solution path that would be found in the absence of any obstacles. The obstacle rotates either $0$ or $\pi/4$ degrees. The obstacle can translate along the x-axis by $0, -0.25, \text{or } 0.25$ m. The horizon length is $N=8$.

We compare the disturbance feedback policies obtained by using three different risk measures - CVaR, EVaR, and TVD by comparing the total cost of reaching the goal, the percentage of infeasible optimizations, and the average computation time for each MPC iteration.

\begin{figure}
    \centering
    \includegraphics[width=100mm]{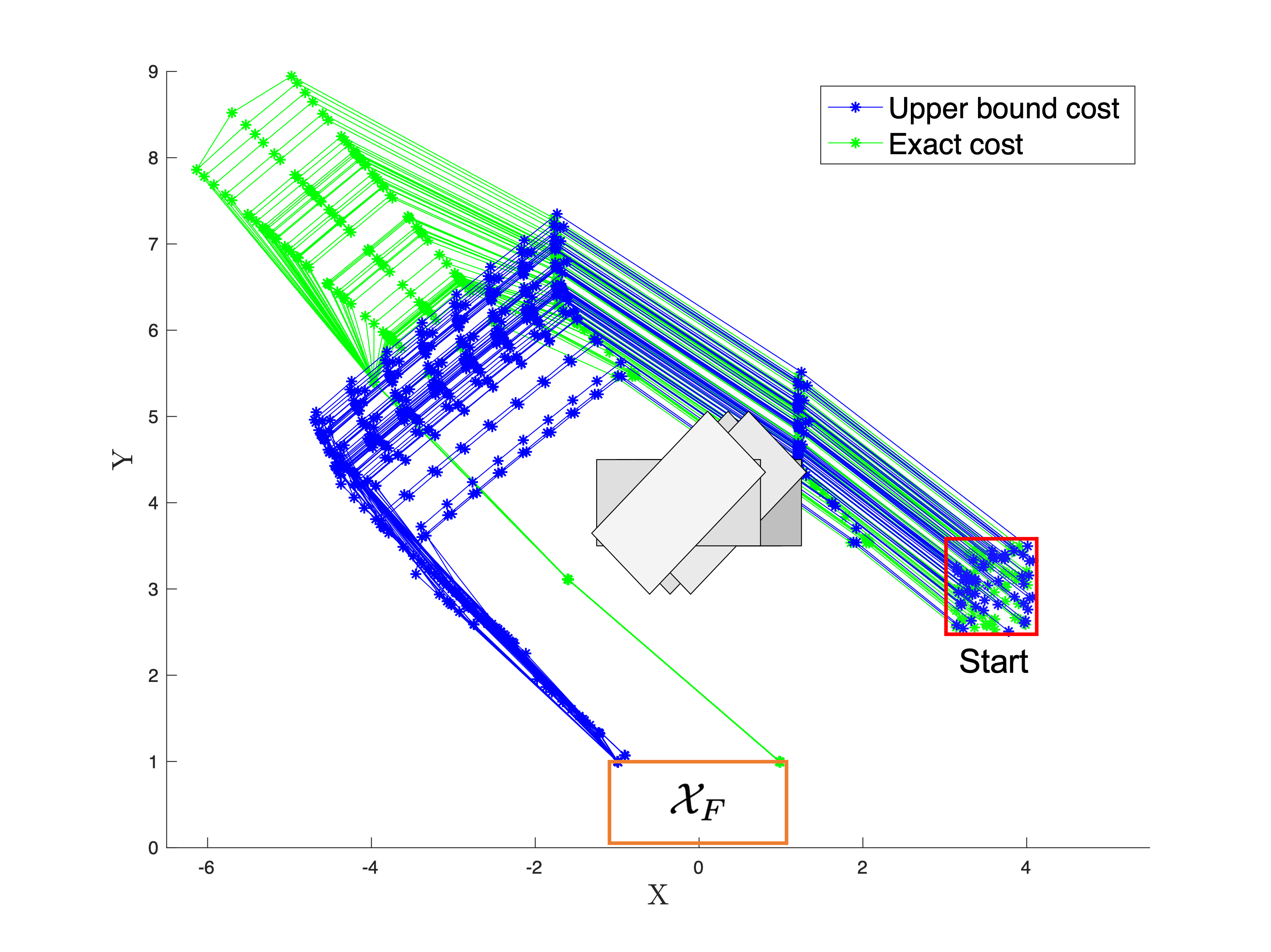}
    \caption{Comparison of the trajectories obtained using the exact cost~\eqref{eq:exact_cost} and the upper bound cost~\eqref{eq:upper_bound_cost}. The gray rectangles show possible obstacle configurations. The darker rectangle has a higher probability of occurrence and the lighter rectangle has a lower probability of occurrence.}
    \label{fig:cost_comp}
\end{figure}

\textbf{Cost Comparison:} A fair comparison of using the exact cost versus the over approximation of the cost can be made only if the constraints of the MPC~\eqref{eq:mpc_exact} remain the same. For 50 Monte Carlo simulations, we compare the MPC trajectories obtained when using CVaR risk. We compare the trajectory costs resulting from using a) the \textit{exact cost} as computed in~\eqref{eq:exact_cost} and b) the \textit{upper bound} of the exact cost~\eqref{eq:upper_bound_cost}, see~\autoref{table:cost_comparison}. The average time taken for each MPC iteration is also provided. We see that the time taken for each MPC iteration is significantly higher when the exact CVaR cost is used. The control effort is also higher when using the exact cost. This is because the over-approximated cost always penalizes higher control effort more than task completion (control is parameterized as an affine function of the disturbance).~\autoref{fig:cost_comp} shows us the qualitative difference between the 50 Monte Carlo simulations when $\alpha = 0.9$. We emphasize that task completion is prioritized when the exact CVaR cost is used. 

\begin{table}[h!]
\centering
\begin{tabular}{||c| c c||} 
 \hline
 $\alpha$ & \multicolumn{2}{c||}{ Average cost ($\times 10^3$)} \\
  &  Exact & Upper bound  \\ [0.5ex]
 \hline\hline
 0.1 & 4.44 & 3.94 \\ 
 \hline
 0.4 & 4.51 &  3.94  \\
 \hline
 0.8 & 5.36 &  3.93  \\
 \hline
 0.9 & 6.20&  3.89 \\ 
 \hline \hline
 Time(s)& 83.68 & 6.32  \\ [1ex] 
 \hline
\end{tabular}
\caption{Average trajectory cost for CVaR MPC using different costs}
\label{table:cost_comparison}
\end{table}


\textbf{Feasibility comparison between different coherent risk measures:} For different risk levels, $\alpha$, we also compare the number of times the MPC optimization is infeasible when we use CVaR, EVaR, and TVD risk measures, with the cost~\eqref{eq:upper_bound_cost}. The results for 50 Monte Carlo simulations are summarized in~\autoref{table:feasibility}. It has been shown that $\text{VaR} \leq \text{CVaR} \leq \text{EVaR}$ and $\text{CVaR} \leq \text{TVD}$, see Section~\ref{section:prelim}. Proposition~\ref{proposition: rec_feasibility} provides us a loose bound for the probability of infeasibility of the MPC.~\autoref{table:feasibility} shows us that as $\alpha$ increases (increasing conservativeness), the percentage of infeasible optimizations decreases. Furthermore, the actual likelihood of infeasibility is much lower than the bounds obtained in Proposition~\ref{proposition: rec_feasibility}. Thus, the bounds of Proposition~\ref{proposition: rec_feasibility} are verified in this case, though the degree of tightness is unknown.

\begin{table}[h!]
\centering
\begin{tabular}{||c | c c c ||} 
 \hline
 $\alpha$ & \multicolumn{3}{c||}{ MPC infeasible (\%)} \\
  &  CVaR & EVaR & TVD\\ [0.5ex]
 \hline\hline
 0.1 & 5.3 & 1.64 & 0\\ 
 \hline
 0.4 & 5.9 & 1.09 & 0 \\
 \hline
 0.8 & 6.4 & 0.18 & 0\\
 \hline
 0.9 & 2.7 & 0 & 0 \\ 
 \hline \hline
 Time (s) & 6.32 & 42.33 & 3.61\\ [1ex] 
 \hline
\end{tabular}
\caption{Results for infeasibility of risk-aware  MPC}
\label{table:feasibility}
\end{table}


\textcolor{black}{\textbf{Comparison with Stochastic MPC for arbitrary distributions:} Popular stochastic MPC techniques like~\cite{blackmore2010particle, blackmore2006probabilistic} are closest to our work. These techniques do not assume Gaussian uncertainty distribution and approximate the joint distribution of the uncertainty using sampling techniques and then provide a chance-constrained particle control algorithm for polytopic obstacle avoidance. For the simple 2D system we've considered here, these techniques are \textit{intractable} without sampling of the joint distribution of uncertainty, i.e., the solution of the MPC cannot be solved within a reasonable time because the number of constraints grow exponentially with horizon length. For the sake of comparison, as discussed in~\cite{blackmore2006probabilistic}, we sample the joint distribution and compare against our TVD-based risk-aware MPC. In our case, we do not have to sample the joint distribution because the constraint-tightening techniques we provide in this paper make the MPC tractable. As discussed in the previous sections, the number of constraints do not grow exponentially with horizon length, but polynomially with the horizon. With fewer number of samples,~\cite{blackmore2010particle, blackmore2006probabilistic} provide approximate, open-loop actions that are less effective in collision avoidance, but the approximate solution is solved faster. We see, in Table~\ref{table:comaprisonSMPC}, as the number of samples grows, so does the time taken to solve the approximate solution. Our method, when using CVaR and TVD (both are MILPs just like SMPC in~\cite{blackmore2010particle,blackmore2006probabilistic}), provides better collision avoidance while solving the  MPC faster than SMPC with 100 samples. }

\textcolor{black}{We emphasize that our MPC technique is robust to a set of distributions, unlike stochastic MPC techniques that require the distribution of the uncertainty to be completely known. In~\cite{dixit2022distributionally}, we further show how TVD-based MPC is effective when the underlying distribution of the process noise is also perturbed.}
\begin{table}[h!]
\centering
\begin{tabular}{||c | c c c c ||} 
 \hline
 & \multicolumn{4}{c||}{ Collision with moving obstacle (\%)} \\
 $\alpha$  &  SMPC (\cite{blackmore2010particle, blackmore2006probabilistic})  & SMPC (\cite{blackmore2010particle, blackmore2006probabilistic}) & CVaR (ours) & TVD (ours)\\
  & (25 samples) & (100 samples) & (No sampling) & (No sampling)\\ [0.5ex]
 \hline\hline
 0.1 & 6 & 4 & 4 &0\\ 
 \hline
 0.9 & 4 & 4 & 0 &2 \\ 
 \hline \hline
 Time (s) & 1.5 & 14.2 & 6.32 &3.61\\[1ex] 
 \hline
\end{tabular}
\caption{\textcolor{black}{Risk-Aware MPC compared against stochastic MPC}}
\label{table:comaprisonSMPC}
\end{table}

\subsection{Quadcopter}

\begin{figure}[tbhp]
\centering
\includegraphics[width=\textwidth]{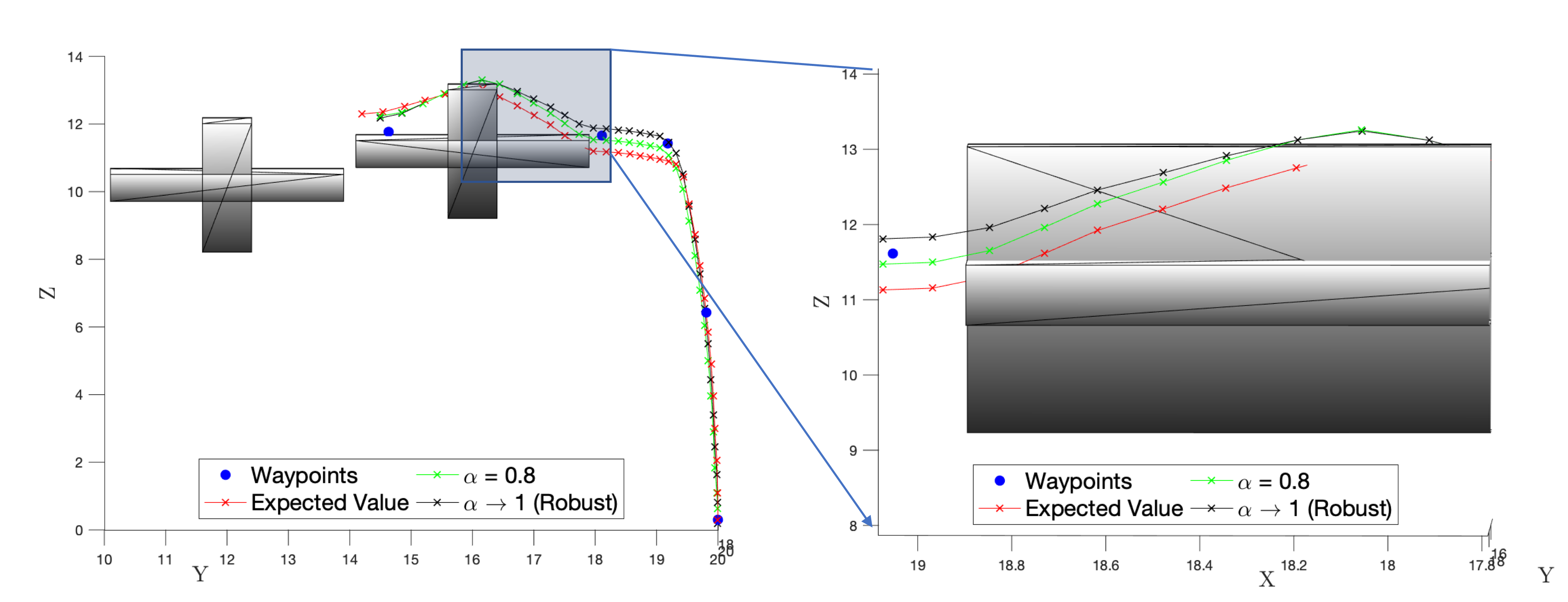}
    \caption{A comparison of the TVD MPC trajectories with the expectation-based MPC ($\alpha\rightarrow0$) trajectory. On the right, the shaded blue region is zoomed in from a different perspective to illustrate the behavior near one of the random realizations of the obstacle.}
    \label{fig:quad_results}
\end{figure}
We consider a quadcopter that must follow given waypoints while avoiding randomly moving obstacles and adhering to state and control constraints, Algorithm~\ref{alg:waypoint_alg}. The quadcopter is described by the states $(x, y, z, \phi, \theta, \varphi, \dot{x}, \dot{y}, \dot{z}, \dot{\phi}, \dot{\theta}, \dot{\varphi})^T$. The position of the quadcopter in 3D space is given by the coordinates $x, y, z$ and the roll, pitch, and yaw are given by $\phi, \theta, \varphi$ respectively. The model of the quadcopter is given by 
    \[ \ddot{x} = -g\theta, \, 
       \ddot{y} = g\theta,\, \ddot{z}=-\frac{u_1}{m} - g, \, \]
    \[ \ddot{\phi} = \frac{u_2}{I_{xx}}, \, 
       \ddot{\theta} = \frac{u_3}{I_{yy}}, \, 
       \ddot{\varphi} = \frac{u_4}{I_{zz}},\]
where $m$ is the quadcopter's mass, $g$ is the acceleration due to gravity, and $I_{xx}, I_{yy}, I_{zz}$ are the quadcopter moments of inertia about the $x,y,z$-axes of the system. The control inputs are given by $u_1, u_2, u_3, u_4$. We used the following parameters: $m = 0.65$kg, $l = 0.23$m, $I_{xx} = 0.0075$kg.m$^2$, $I_{yy} = 0.0075$kg.m$^2$, $I_{zz} = 0.0013$kg.m$^2$, $g = 9.81$m.s$^{-2}$~\cite{dixit2020risksensitive, hakobyan2019cvar}.

The TVD risk constraint has two parameters: the confidence level, $\alpha$, and the risk-threshold, $\epsilon$. We chose $\alpha \in \{0.8, 0.99\},\, \epsilon = 0.04$. The waypoints are given by regions of size $[-0.5,0.5]^3$ around the waypoint center (denoted by o in~\autoref{fig:quad_results}). We chose a horizon length of $N = 15$ for the MPC optimization. We considered the case of having one randomly translating and rotating obstacle. The obstacle is a rectangular box of size $2$x$1$x$4$ m$^3$; it can rotate by $\frac{\pi}{2}$ and translate by $4$m along the y-axis and $1$m along the z-axis.~\autoref{fig:quad_results} shows all the different configurations of this obstacle at different times. We further allow for process noise $\delta \in \{ -0.2, -0.1, 0, 0.1, 0.2\}$. The continuous-time system is discretized with a time-interval of $0.02$ sec. \textcolor{black}{The average computation time for each iteration is 5.4 seconds.} As the risk-averseness of the system grows, the trajectory followed by the quadcopter maintains a greater distance from all possible configurations of the moving obstacle.

\section{Conclusions}~\label{sec:conclusions}
This work proposed a risk-aware framework for motion planning with obstacle avoidance using MPC. We provided reformulations of the risk constraint and the cost to obtain a tractable, convex, mixed-integer optimization. We also provided guarantees on the recursive feasibility and the finite-time task completion of the MPC. Coherent risk measures make the system distributionally robust to disturbances and this provides stronger guarantees on system safety compared to expectation-based MPC (or stochastic MPC). This risk-aware formulation is a generalization of stochastic MPC and robust MPC into one framework through the adjustment of the risk-level $\alpha$. \textcolor{black}{We showed through numerical experiments that our framework can provide a tractable solution to the exact obstacle avoidance problem while stochastic MPC methods need sampling techniques to be able to provide a tractable solution with approximate obstacle avoidance. The comparison between our method and SMPC for a simple 2D system showed better obstacle avoidance with solve times comparable to sampling-based SMPC. }

There are several ways in which this formulation can be improved. As seen in the numerical results, the time taken for each MPC optimization, is a few seconds. This does not allow for real-time application of the MPC. The MPC can be sped-up by removing (or reducing the number of) the mixed-integer constraints. For the obstacle avoidance constraint, one can find half-space approximations of the obstacle~\cite{wei2022moving,nair2021stochastic}. The formulation also accounts for only discrete probability distributions. An extension to continuous distributions will require approximation methods like Sample Average Approximation~\cite{kim2015guide}. Furthermore, coherent risk measures provide distributional robusteness in the expectation of the cost. A natural extension is to incorporate distributionally robust chance constraints in the MPC~\cite{dixit2022distributionally}. Lastly, this work is limited to linear, discrete-time systems. Future work involves accounting for nonlinear dynamics and providing continuous-time safety guarantees in a risk-aware manner~\cite{jasour2021convex}.  


\section*{Acknowledgement}
This work was funded in part by DARPA, through the Subterranean Challenge program. The authors would like to thank Skylar Wei and Thomas Touma for helping run many simulations. The authors would also like to acknowledge valuable input from Prithvi Akella, Skylar Wei, and Siddharth Nair. \textcolor{black}{The authors would also like to thank the reviewers for their feedback that has helped improved the paper significantly.}
\balance

\bibliography{IEEEexample}

\begin{thebibliography}{10}
\expandafter\ifx\csname url\endcsname\relax
  \def\url#1{\texttt{#1}}\fi
\expandafter\ifx\csname urlprefix\endcsname\relax\def\urlprefix{URL }\fi
\expandafter\ifx\csname href\endcsname\relax
  \def\href#1#2{#2} \def\path#1{#1}\fi

\bibitem{bouman2020autonomous}
A.~Bouman, M.~F. Ginting, N.~Alatur, M.~Palieri, D.~D. Fan, T.~Touma,
  T.~Pailevanian, S.-K. Kim, K.~Otsu, J.~Burdick, et~al., {Autonomous Spot:
  Long-Range Autonomous Exploration of Extreme Environments with Legged
  Locomotion}, arXiv preprint arXiv:2010.09259.

\bibitem{fan2021step}
D.~D. Fan, K.~Otsu, Y.~Kubo, A.~Dixit, J.~Burdick, A.-A. Agha-Mohammadi, Step:
  Stochastic traversability evaluation and planning for risk-aware off-road
  navigation, in: Robotics: Science and Systems, RSS Foundation, 2021, pp.
  1--21.

\bibitem{daftry2022mlnav}
S.~Daftry, N.~Abcouwer, T.~D. Sesto, S.~Venkatraman, J.~Song, L.~Igel, A.~Byon,
  U.~Rosolia, Y.~Yue, M.~Ono, Mlnav: Learning to safely navigate on martian
  terrains, IEEE Robotics and Automation Letters 7~(2) (2022) 5461--5468.
\newblock \href {http://dx.doi.org/10.1109/LRA.2022.3156654}
  {\path{doi:10.1109/LRA.2022.3156654}}.

\bibitem{nagatani2013emergency}
K.~Nagatani, S.~Kiribayashi, Y.~Okada, K.~Otake, K.~Yoshida, S.~Tadokoro,
  T.~Nishimura, T.~Yoshida, E.~Koyanagi, M.~Fukushima, et~al., Emergency
  response to the nuclear accident at the fukushima daiichi nuclear power
  plants using mobile rescue robots, Journal of Field Robotics 30~(1) (2013)
  44--63.

\bibitem{seraj2020coordinated}
E.~Seraj, M.~Gombolay, Coordinated control of uavs for human-centered active
  sensing of wildfires, in: 2020 American Control Conference (ACC), IEEE, 2020,
  pp. 1845--1852.

\bibitem{rosolia2017lmpc}
U.~Rosolia, A.~Carvalho, F.~Borrelli, Autonomous racing using learning model
  predictive control, in: 2017 American Control Conference (ACC), 2017, pp.
  5115--5120.
\newblock \href {http://dx.doi.org/10.23919/ACC.2017.7963748}
  {\path{doi:10.23919/ACC.2017.7963748}}.

\bibitem{rosolia2020unified}
U.~Rosolia, A.~Singletary, A.~D. Ames, Unified multi-rate control: from low
  level actuation to high level planning, arXiv preprint arXiv:2012.06558.

\bibitem{ames2017cbf}
A.~D. Ames, X.~Xu, J.~W. Grizzle, P.~Tabuada, Control barrier function based
  quadratic programs for safety critical systems, IEEE Transactions on
  Automatic Control 62~(8) (2017) 3861--3876.
\newblock \href {http://dx.doi.org/10.1109/TAC.2016.2638961}
  {\path{doi:10.1109/TAC.2016.2638961}}.

\bibitem{blackmore2011CC}
L.~Blackmore, M.~Ono, B.~C. Williams, Chance-constrained optimal path planning
  with obstacles, IEEE Transactions on Robotics 27~(6) (2011) 1080--1094.
\newblock \href {http://dx.doi.org/10.1109/TRO.2011.2161160}
  {\path{doi:10.1109/TRO.2011.2161160}}.

\bibitem{jasourrisk}
A.~M. Jasour, B.~C. Williams, Risk contours map for risk bounded motion
  planning under perception uncertainties, Robotics: Science and Systems.

\bibitem{luders2010chance}
B.~Luders, M.~Kothari, J.~How, Chance constrained rrt for probabilistic
  robustness to environmental uncertainty, in: AIAA guidance, navigation, and
  control conference, 2010, p. 8160.

\bibitem{Aoude2013ProbabilisticallySM}
G.~Aoude, B.~Luders, J.~M. Joseph, N.~Roy, J.~P. How, Probabilistically safe
  motion planning to avoid dynamic obstacles with uncertain motion patterns,
  Autonomous Robots 35 (2013) 51--76.

\bibitem{lindemann2021robust}
L.~Lindemann, M.~Cleaveland, Y.~Kantaros, G.~J. Pappas,
  \href{https://arxiv.org/abs/2108.11983}{Robust motion planning in the
  presence of estimation uncertainty} (2021).
\newblock \href {http://dx.doi.org/10.48550/ARXIV.2108.11983}
  {\path{doi:10.48550/ARXIV.2108.11983}}.
\newline\urlprefix\url{https://arxiv.org/abs/2108.11983}

\bibitem{kim2021plgrim}
S.-K. Kim, A.~Bouman, G.~Salhotra, D.~D. Fan, K.~Otsu, J.~Burdick, A.-a.
  Agha-mohammadi, Plgrim: Hierarchical value learning for large-scale
  exploration in unknown environments, in: Proceedings of the International
  Conference on Automated Planning and Scheduling, Vol.~31, 2021, pp. 652--662.

\bibitem{borrelli2003constrained}
F.~Borrelli, Constrained optimal control of linear and hybrid systems, Vol.
  290, Springer, 2003.

\bibitem{nair2021stochastic}
S.~H. Nair, V.~Govindarajan, T.~Lin, C.~Meissen, H.~E. Tseng, F.~Borrelli,
  Stochastic mpc with multi-modal predictions for traffic intersections, arXiv
  preprint arXiv:2109.09792.

\bibitem{bemporad1999robust}
A.~Bemporad, M.~Morari, Robust model predictive control: A survey, in:
  Robustness in identification and control, Springer, 1999, pp. 207--226.

\bibitem{mesbah2016SMPC}
A.~Mesbah, Stochastic model predictive control: An overview and perspectives
  for future research, IEEE Control Systems Magazine 36~(6) (2016) 30--44.
\newblock \href {http://dx.doi.org/10.1109/MCS.2016.2602087}
  {\path{doi:10.1109/MCS.2016.2602087}}.

\bibitem{jasour2019risk}
A.~Jasour, Risk aware and robust nonlinear planning (rarnop), Course Notes for
  MIT 16 (2019) S498.

\bibitem{ono2015chance}
M.~Ono, M.~Pavone, Y.~Kuwata, J.~Balaram, Chance-constrained dynamic
  programming with application to risk-aware robotic space exploration,
  Autonomous Robots 39~(4) (2015) 555--571.

\bibitem{han2022non}
W.~Han, A.~Jasour, B.~Williams, Non-gaussian risk bounded trajectory
  optimization for stochastic nonlinear systems in uncertain environments,
  arXiv preprint arXiv:2203.03038.

\bibitem{Hyeon2020fast}
E.~Hyeon, Y.~Kim, A.~G. Stefanopoulou, Fast risk-sensitive model predictive
  control for systems with time-series forecasting uncertainties, in: 2020 59th
  IEEE Conference on Decision and Control (CDC), 2020, pp. 2515--2520.
\newblock \href {http://dx.doi.org/10.1109/CDC42340.2020.9304447}
  {\path{doi:10.1109/CDC42340.2020.9304447}}.

\bibitem{moehle2021risk}
N.~Moehle, Risk-sensitive model predictive control, arXiv preprint
  arXiv:2101.11166.

\bibitem{koenig1994risk}
S.~Koenig, R.~G. Simmons, Risk-sensitive planning with probabilistic decision
  graphs, in: Principles of Knowledge Representation and Reasoning, Elsevier,
  1994, pp. 363--373.

\bibitem{xu2010distributionally}
H.~Xu, S.~Mannor, {Distributionally robust Markov decision processes}, in:
  Advances in Neural Information Processing Systems, 2010, pp. 2505--2513.

\bibitem{chen2022distributionally}
Y.~Chen, J.~Kim, J.~Anderson, Distributionally robust decision making
  leveraging conditional distributions, arXiv preprint arXiv:2204.00138.

\bibitem{RENGANATHAN202015530}
V.~Renganathan, I.~Shames, T.~H. Summers,
  \href{https://www.sciencedirect.com/science/article/pii/S2405896320330780}{Towards
  integrated perception and motion planning with distributionally robust risk
  constraints⁎⁎this work is partially supported by defence science and
  technology group, through agreement myip: Id9156 entitled “verifiable
  hierarchical sensing, planning and control”, the australian government, via
  grant ausmurib000001 associated with onr muri grant n00014-19-1-2571, and by
  the united states air force office of scientific research under award number
  fa2386-19-1-4073.}, IFAC-PapersOnLine 53~(2) (2020) 15530--15536, 21st IFAC
  World Congress.
\newblock \href
  {http://dx.doi.org/https://doi.org/10.1016/j.ifacol.2020.12.2396}
  {\path{doi:https://doi.org/10.1016/j.ifacol.2020.12.2396}}.
\newline\urlprefix\url{https://www.sciencedirect.com/science/article/pii/S2405896320330780}

\bibitem{nair2022collision}
S.~H. Nair, E.~H. Tseng, F.~Borrelli, Collision avoidance for dynamic obstacles
  with uncertain predictions using model predictive control, in: 2022 IEEE 61st
  Conference on Decision and Control (CDC), 2022, pp. 5267--5272.
\newblock \href {http://dx.doi.org/10.1109/CDC51059.2022.9993319}
  {\path{doi:10.1109/CDC51059.2022.9993319}}.

\bibitem{majumdar2020should}
A.~Majumdar, M.~Pavone, How should a robot assess risk? towards an axiomatic
  theory of risk in robotics, in: Robotics Research, Springer, 2020, pp.
  75--84.

\bibitem{artzner1999coherent}
P.~Artzner, F.~Delbaen, J.-M. Eber, D.~Heath, Coherent measures of risk,
  Mathematical finance 9~(3) (1999) 203--228.

\bibitem{singh2018framework}
S.~Singh, Y.~Chow, A.~Majumdar, M.~Pavone, A framework for time-consistent,
  risk-sensitive model predictive control: Theory and algorithms, IEEE
  Transactions on Automatic Control.

\bibitem{wang2021adaptive}
Z.~Wang, O.~So, K.~Lee, E.~A. Theodorou, Adaptive risk sensitive model
  predictive control with stochastic search, in: Learning for Dynamics and
  Control, PMLR, 2021, pp. 510--522.

\bibitem{hakobyan2019cvar}
A.~{Hakobyan}, G.~C. {Kim}, I.~{Yang}, Risk-aware motion planning and control
  using cvar-constrained optimization, IEEE Robotics and Automation Letters
  4~(4) (2019) 3924--3931.

\bibitem{dixit2020risksensitive}
A.~Dixit, M.~Ahmadi, J.~W. Burdick, Risk-sensitive motion planning using
  entropic value-at-risk, in: European Control Conference, 2021.

\bibitem{ahmadi2020cvar}
M.~Ahmadi, X.~Xiong, A.~D. Ames, Risk-averse control via {CVaR} barrier
  functions: Application to bipedal robot locomotion, IEEE Control Systems
  Letters 6 (2021) 878--883.

\bibitem{Sopasakis2019}
P.~{Sopasakis}, M.~{Schuurmans}, P.~{Patrinos}, Risk-averse risk-constrained
  optimal control, in: 2019 18th European Control Conference (ECC), 2019, pp.
  375--380.
\newblock \href {http://dx.doi.org/10.23919/ECC.2019.8796021}
  {\path{doi:10.23919/ECC.2019.8796021}}.

\bibitem{chen2022interactive}
Y.~Chen, U.~Rosolia, W.~Ubellacker, N.~Csomay-Shanklin, A.~D. Ames, Interactive
  multi-modal motion planning with branch model predictive control, IEEE
  Robotics and Automation Letters 7~(2) (2022) 5365--5372.

\bibitem{SCHUURMANS202015128}
M.~Schuurmans, A.~Katriniok, H.~E. Tseng, P.~Patrinos, Learning-based
  risk-averse model predictive control for adaptive cruise control with
  stochastic driver models, IFAC-PapersOnLine 53~(2) (2020) 15128--15133, 21st
  IFAC World Congress.
\newblock \href
  {http://dx.doi.org/https://doi.org/10.1016/j.ifacol.2020.12.2037}
  {\path{doi:https://doi.org/10.1016/j.ifacol.2020.12.2037}}.

\bibitem{ahmadi2012entropic}
A.~Ahmadi-Javid, Entropic value-at-risk: A new coherent risk measure, Journal
  of Optimization Theory and Applications 155~(3) (2012) 1105--1123.

\bibitem{shapiro2017distributionally}
A.~Shapiro, Distributionally robust stochastic programming, SIAM Journal on
  Optimization 27~(4) (2017) 2258--2275.

\bibitem{NAVRATIL198863}
J.~Navratil, K.~Lim, D.~Fisher, Disturbance feedback in model predictive
  control systems, IFAC Proceedings Volumes 21~(4) (1988) 63--68, iFAC Workshop
  on Model Based Process Control, Atlanta, GA, USA, 13-14 June.
\newblock \href
  {http://dx.doi.org/https://doi.org/10.1016/B978-0-08-035735-5.50013-9}
  {\path{doi:https://doi.org/10.1016/B978-0-08-035735-5.50013-9}}.

\bibitem{Oldewurtel2008affineDF}
F.~Oldewurtel, C.~N. Jones, M.~Morari, A tractable approximation of chance
  constrained stochastic mpc based on affine disturbance feedback, in: 2008
  47th IEEE Conference on Decision and Control, 2008, pp. 4731--4736.
\newblock \href {http://dx.doi.org/10.1109/CDC.2008.4738806}
  {\path{doi:10.1109/CDC.2008.4738806}}.

\bibitem{GOULART2006523}
P.~J. Goulart, E.~C. Kerrigan, J.~M. Maciejowski,
  \href{https://www.sciencedirect.com/science/article/pii/S0005109806000021}{Optimization
  over state feedback policies for robust control with constraints}, Automatica
  42~(4) (2006) 523--533.
\newblock \href
  {http://dx.doi.org/https://doi.org/10.1016/j.automatica.2005.08.023}
  {\path{doi:https://doi.org/10.1016/j.automatica.2005.08.023}}.
\newline\urlprefix\url{https://www.sciencedirect.com/science/article/pii/S0005109806000021}

\bibitem{zhang2021SADF}
J.~Zhang, T.~Ohtsuka, Stochastic model predictive control using simplified
  affine disturbance feedback for chance-constrained systems, IEEE Control
  Systems Letters 5~(5) (2021) 1633--1638.
\newblock \href {http://dx.doi.org/10.1109/LCSYS.2020.3042085}
  {\path{doi:10.1109/LCSYS.2020.3042085}}.

\bibitem{bertsekas2009convex}
D.~Bertsekas, Convex optimization theory, Vol.~1, Athena Scientific, 2009.

\bibitem{vecchietti_modeling_2003}
A.~Vecchietti, S.~Lee, I.~Grossman, Modeling of {Discrete}/{Continuous}
  {Optimization} {Problems}: {Characterization} and {Formulation} of
  {Disjunctions} and {Their} {Relaxations}, Computers and Chemical Engineering
  27~(3) (2003) 433--448.

\bibitem{richards2003robustFeasibility}
A.~{Richards}, J.~P. {How}, Model predictive control of vehicle maneuvers with
  guaranteed completion time and robust feasibility, in: Proceedings of the
  2003 American Control Conference, 2003., Vol.~5, 2003, pp. 4034--4040 vol.5.
\newblock \href {http://dx.doi.org/10.1109/ACC.2003.1240467}
  {\path{doi:10.1109/ACC.2003.1240467}}.

\bibitem{Lopez2017}
B.~T. {Lopez}, J.~P. {How}, Aggressive collision avoidance with limited
  field-of-view sensing, in: 2017 IEEE/RSJ International Conference on
  Intelligent Robots and Systems (IROS), 2017, pp. 1358--1365.
\newblock \href {http://dx.doi.org/10.1109/IROS.2017.8202314}
  {\path{doi:10.1109/IROS.2017.8202314}}.

\bibitem{yalmip}
J.~Lofberg, Yalmip: A toolbox for modeling and optimization in matlab, in:
  Computer Aided Control Systems Design, 2004 IEEE International Symposium on,
  IEEE, 2004, pp. 284--289.

\bibitem{gurobi}
{Gurobi Optimization, LLC}, \href{https://www.gurobi.com}{{Gurobi Optimizer
  Ref. Manual}} (2022).
\newline\urlprefix\url{https://www.gurobi.com}

\bibitem{mosek}
M.~ApS, \href{http://docs.mosek.com/9.0/toolbox/index.html}{The MOSEK
  optimization toolbox for MATLAB manual. Version 9.0.} (2019).
\newline\urlprefix\url{http://docs.mosek.com/9.0/toolbox/index.html}

\bibitem{blackmore2010particle}
L.~Blackmore, M.~Ono, A.~Bektassov, B.~C. Williams, A probabilistic
  particle-control approximation of chance-constrained stochastic predictive
  control, IEEE Transactions on Robotics 26~(3) (2010) 502--517.
\newblock \href {http://dx.doi.org/10.1109/TRO.2010.2044948}
  {\path{doi:10.1109/TRO.2010.2044948}}.

\bibitem{blackmore2006probabilistic}
L.~Blackmore, A probabilistic particle control approach to optimal, robust
  predictive control, in: AIAA Guidance, Navigation, and Control Conference and
  Exhibit, 2006, p. 6240.

\bibitem{dixit2022distributionally}
A.~Dixit, M.~Ahmadi, J.~W. Burdick, Distributionally robust model predictive
  control with total variation distance, IEEE Control Systems Letters 6 (2022)
  3325--3330.
\newblock \href {http://dx.doi.org/10.1109/LCSYS.2022.3184921}
  {\path{doi:10.1109/LCSYS.2022.3184921}}.

\bibitem{wei2022moving}
S.~X. Wei, A.~Dixit, S.~Tomar, J.~W. Burdick, Moving obstacle avoidance: A
  data-driven risk-aware approach, IEEE Control Systems Letters 7 (2022)
  289--294.
\newblock \href {http://dx.doi.org/10.1109/LCSYS.2022.3181191}
  {\path{doi:10.1109/LCSYS.2022.3181191}}.

\bibitem{kim2015guide}
S.~Kim, R.~Pasupathy, S.~G. Henderson, A guide to sample average approximation,
  Handbook of simulation optimization (2015) 207--243.

\bibitem{jasour2021convex}
A.~Jasour, W.~Han, B.~Williams, Convex risk bounded continuous-time trajectory
  planning in uncertain nonconvex environments, Robotics: Science and Systems.

\bibitem{2004boyd}
S.~Boyd, L.~Vandenberghe, Convex Optimization, Cambridge University Press, USA,
  2004.

\end{thebibliography}

\appendix
\section{Proof of Lemma~\ref{lem:stcon_tighten}}
\begin{proof}
First, rewrite the risk state constraint~\eqref{eq:state_const} as,
\begin{align*}
    \rho(f_{x,n}^T(A^{k}x_{0} + \boldsymbol{B}_k\boldsymbol{u}_{k} + \boldsymbol{D}_k\boldsymbol{\delta}_{k}) - g_{x,n}) \leq \epsilon_x, \quad  \forall  k\in\mathbb{Z}_1^{N}, n\in\mathbb{Z}_1^{r}
\end{align*}
where, $\boldsymbol{B}_k = \begin{bmatrix}A^{k-1}B & A^{k-2}B & \dotsc & B\end{bmatrix}$, and $\boldsymbol{D}_k = \begin{bmatrix}A^{k-1}D & A^{k-2}D & \dotsc & D\end{bmatrix}$. 
We replace $\boldsymbol{u}_{k}$ with the SADF control policy~\eqref{eq:controlLaw1},
\begin{equation}\label{eq: state_constraint_batch}
\begin{aligned}
\epsilon_x &\geq
 \rho\big(f_{x,n}^T\big(A^{k}x_{0} + \boldsymbol{B}_k(\boldsymbol{\eta}_{k} + \boldsymbol{K}_k\boldsymbol{\delta}_{k}) + \boldsymbol{D}_k\boldsymbol{\delta}_{k}\big) - g_{x,n}\big) \\
    & = f_{x,n}^T\big(A^{k}x_{0} + \boldsymbol{B}_k\boldsymbol{\eta}_{k}\big) + \rho\big(f_{x,n}^T(\boldsymbol{B}_k\boldsymbol{K}_k + \boldsymbol{D}_k)\boldsymbol{\delta}_{k}\big) - g_{x,n} \\
    &=f_{x,n}^T\big(A^{k}x_{0} + \boldsymbol{B}_k\boldsymbol{\eta}_{k}\big) +\rho\big(f_{x,n}^T\sum_{m=0}^{k-1}\big(A^{k-m-1}(BK_{k-m} + D)\big){\delta}_{m}\big) -g_{x,n} 
\end{aligned}
\end{equation}    
\normalsize The second term on the right-hand side of the above inequality can be simplified by using the subadditivity, monotonicity, and positive homogeneity properties of coherent risk measures and then i.i.d. nature of the disturbances respectively,  
\begin{align*}
     &\rho\bigg(f_{x,n}^T\sum_{m=0}^{k-1}\big(A^{k-m-1}(BK_{k-m} + D)\big){\delta}_{m}\bigg) &\\&\leq  \sum_{m=0}^{k-1}\rho\bigg(f_{x,n}^T\big(A^{k-m-1}(BK_{k-m} + D)\big){\delta}_{m}\bigg)  & \text{(Subadditivity)} \\ &\leq
    \sum_{m=0}^{k-1}\rho\bigg(|f_{x,n}^T\big(A^{k-m-1}(BK_{k-m} + D)\big)|\,|{\delta}_{m}|\bigg) & \text{(Monotonicity)} \\ &\leq
    \sum_{m=0}^{k-1}|f_{x,n}^T\big(A^{k-m-1}(BK_{k-m} + D)\big)|\,\rho(|{\delta}_{m}|) & \text{(Positive Homogeneity)} \\ &\leq
    \lVert f_{x,n}^T\big(\boldsymbol{B}_k\boldsymbol{K}_k + \boldsymbol{D}_k\big)\rVert_{1}\rho(|{\delta}|). & \text{(i.i.d disturbances)}
\end{align*}
\normalsize Hence, satisfying the tightened constraint \eqref{eq:state_tighten} implies satisfaction of the state constraint~\eqref{eq:state_const}.
\end{proof}

\section{Proof of Lemma~\ref{lem: safety_constraint}}
\begin{proof}
To find the distance of $\boldsymbol{y}_k$ from the safe set, $\zeta(\boldsymbol{y}_k, \mathcal{S}_l(t+k))$, we define a set of variables $h_{l,k}^j\ge 0$, $l \in \mathbb{Z}_1^L$ and $k=0,\ldots,N-1$ satisfying
\begin{equation}\label{eq:safeset1}
    \boldsymbol{y}_k +\frac{\boldsymbol{c}_{i,l}}{||\boldsymbol{c}_{i,l}||}h_{l,k}^j = \boldsymbol{z}
\end{equation}
 $\forall j \in \mathbb{Z}_1^J, \forall k \in \mathbb{Z}_0^{N-1}$ and for some $i \in \mathbb{Z}_1^{m_l}$, which is the distance from each $\boldsymbol{y}_k$ to a point $\boldsymbol{z} \in \mathcal{X}$. If $\boldsymbol{z} \in \mathcal{S}_l^j(t+k)$, then  minimizing $h_{l,k} $ defines the line segment connecting $ \boldsymbol{y}_k$ and $\boldsymbol{z}$, which is the minimum distance to the set $ \mathcal{S}_l^j(t+k)$. Therefore,

 \begin{equation}\label{eq:safeset1}
 \begin{split}
    \zeta(\boldsymbol{y}_k, \mathcal{S}_l(t+k)) &= \min_{\boldsymbol{z} \in \mathcal{S}_l(t+k)} ||\boldsymbol{y}(t+k) - \boldsymbol{z}||\\ &= \left \lbrace\begin{matrix} \min_{h_{l,k}} & h_{l,k} \qquad \\
     \text{s.t.}&\boldsymbol{y}_k +\frac{\boldsymbol{c}_{i,l}}{||\boldsymbol{c}_{i,l}||}h_{l,k}^j \in \mathcal{S}_l^j(t+k), 
    \end{matrix} \right.
    \end{split}
 \end{equation}
 and we denote $h_{l,k}^*$ as the solution to~\eqref{eq:safeset1}.
 
Next, substitute the dual form of coherent risk measures (given by the representation theorem) from \eqref{eq:evar} into the L.H.S. of \eqref{eq:safetycon}. Then, replace the risk envelope $\mathcal{Q}$ with the convex representation given in Assumption~\ref{assumption: riskseparable}. That is, $\rho(\zeta(\boldsymbol{y}_k, \mathcal{S}_l(t+k))) = \max_{Q\in\mathcal{Q}}\mathbb{E}_Q\big[\zeta(\boldsymbol{y}_k, \mathcal{S}_l(t+k))\big]=\max_{Q\in\mathcal{Q}}\mathbb{E}_Q\big[h_{l,k}^*\big]$, where~\eqref{eq:safeset1} is used.
\begin{equation}\label{eq:dual}
    \begin{aligned}
        \max_{q(1), \dotsc, q(J)} \quad & \mathbb{E}_Q\big[h_{l,k}^{j*}\big] & \\
        \text{s.t.} \quad & {g}(q) \leq 0, \, -q(j)\leq 0, & \forall j\in\mathbb{Z}_1^J, \\
        &\sum_{j\in\mathbb{Z}_1^J}p(j)q(j) = 1. & 
    \end{aligned}
\end{equation}

The dual of this problem is given by,
\begin{equation}\label{eq:dualofdual}
\begin{aligned}
      \min_{\boldsymbol{\lambda}_1, {\lambda}_2, \nu} \max_{\boldsymbol{q}} & { \big\{\sum_{j\in\mathbb{Z}_1^J}\big[q(j)p(j)h_{l,k}^{j,*} + \lambda^j_1q(j) + \nu p(j)q(j)\big]
    -{\lambda}_2{g}(q) - \nu \big\}}\\
    \text{s.t.}\qquad& \boldsymbol{\lambda}_1 \succeq 0,\, {\lambda}_2 \geq 0, 
\end{aligned}
\end{equation}
where, $\boldsymbol{\lambda}_1 = \begin{bmatrix} \lambda^1_1, \dotsc, \lambda^{J}_1\end{bmatrix}\in\mathbb{R}^{J}, \, {\lambda}_2 \text{ and } \nu \in \mathbb{R}$ are the dual variables.
We conclude that \eqref{eq:dualofdual} and \eqref{eq:dual} are equivalent because strong duality holds by Slater's condition \cite{2004boyd}. Slater's condition is satisfied by showing \textit{strict feasibility} for \eqref{eq:dual}, i.e., there exists a feasible solution to \eqref{eq:dual} such that the inequality constraints hold with strict inequalities. One such solution occurs when $q(j) = 1,\, \forall j\in \mathbb{Z}_1^J$. We can find the maximum value of the Lagrangian in the objective of~\eqref{eq:dualofdual} when we know the exact form of the function $g$.

We can equivalently write the inner maximization of \eqref{eq:dualofdual} in the form of the convex conjugate of $g$ given by $g^*$
\begin{equation}\label{eq:conjugate}
\begin{aligned}
    \min_{\boldsymbol{\lambda}_1, \lambda_2, \nu} \quad& \lambda_2g^{*}\bigg(\lambda_2^{-1}\big(\boldsymbol{p}(h_{l,k}^{j,*} +\nu)+\boldsymbol{\lambda}_1\big) \bigg)-\nu&\\
    \text{s.t.}\quad& \boldsymbol{\lambda}_1 \succeq 0,\, \lambda_2 \geq 0  & \\
    &\lambda_2^{-1}\big(\boldsymbol{p}(h_{l,k}^{j,*} +\nu)+\boldsymbol{\lambda}_1\big) \in \mathbb{R}^J.&
    \end{aligned}
\end{equation}
The above minimization is convex in the dual variables because the perspective operation preserves convexity \cite{2004boyd}. We know from the conjugacy theorem (\cite{bertsekas2009convex}, Proposition 1.6.1) that properness of $g$ implies properness of $g^{*}$. \textcolor{black}{The domain of $g^*$ is the dual space of the domain of $g$, i.e., $\lambda_2^{-1}\big(\boldsymbol{p}(h_{l,k}^{j,*} +\nu)+\boldsymbol{\lambda}_1\big) \in \mathbb{R}^J$, which implies that $\lambda_2\neq 0$.}
Finally, substituting minimization problem~\eqref{eq:safeset1} for $h^*_{l,k}$ in  optimization \eqref{eq:conjugate} gives  \eqref{eq: safetycon_reformulate}. 
\end{proof}
\begin{rmk}
The term $\boldsymbol{p}(h_{l,k}^* +\nu)$ in~\eqref{eq:conjugate} is just the expected value, i.e., $\boldsymbol{p}(h_{l,k}^* +\nu) = \mathbb{E}_P(h_{l,k}^* +\nu)$.
\end{rmk}
\section{Proof of Proposition~\ref{proposition: rec_feasibility}}
\begin{proof}
 Assume that the feasible solution to \eqref{eq:mpc_exact} at time $t$ is given by the control policy $\{(0, \eta_0^*),({K}_1^*, \eta_1^*),\dotsc , ({K}_{N-1}^*,\eta_{N-1}^*)\}$. At time $t$, we apply the control input $u_0 = \eta_0^*$. Since~\eqref{eq:state_tighten},~\eqref{eq:control_tighten} hold for all $\delta_0 \leq \rho(|\delta|)$, if $\delta_0 > \rho(|\delta|)$, the state and control constraints may not hold and the MPC optimization may no longer be feasible. Similarly, if the distance to the obstacle is greater than the risk of the distance, the MPC may no longer be feasible, i.e., if $h_{l,0} > \rho(h_{l,0})$. 
 
Let us assume for simplicity that $\delta_0 \leq \rho(|\delta|),\, h_{l,0} \leq \rho(h_{l,0})$ hold at time $t$. The optimization is feasible at time $t+1$ if there exists a feasible input at time $t+N$ that does not violate constraints. Since $\psi_N^* = 0$ by virtue of the terminal constraint, all the constraints in the optimization are relaxed thereafter. Note that the state $\psi_N = 0$ is invariant due to \eqref{eq:taskCompletion} and \eqref{eq:discreteStateConst} and that $\mu_k = 0$ at all time after the task has been completed. Therefore, once the optimization constraints are relaxed they will remain this way. A control input $(K_N, \eta_N)= (0, 0)$ ensures that the optimization is feasible. At time $t+1$, a feasible solution to \eqref{eq:mpc_exact} is given by the control sequence $\{(0, {K}_1^*\delta_0+\eta_1^*),\dotsc , ({K}_{N-1}^*,\eta_{N-1}^*), (0, 0)\}$.
 
 Finally, we aim to quantify the probability of the constraints at time $t+1$ no longer being satisfied by the control input $u_0 = \eta_0^*$.
 \begin{align*}
      \mathbb{P}\{&\text{MPC infeasible}\} \\
      &\leq \mathbb{P}\{\delta_0 > \rho(|\delta|) \, \cup\, h_{l,0} > \rho(h_{l,0})\} &\\
      &\leq  \mathbb{P}\{\delta_0 \geq \rho(|\delta|)\} + \mathbb{P}\{h_{l,0} \geq \rho(h_{l,0})\} &\\
      &= \mathbb{P}\{\delta_0 - \mathbb{E}(|\delta|) \geq \rho(|\delta|)- \mathbb{E}(|\delta|)\} + \mathbb{P}\{h_{l,0} - \mathbb{E}(h_{l,0})\geq \rho(h_{l,0})- \mathbb{E}(h_{l,0})\}& \\& { \text{(Subtracting $\mathbb{E}(|\delta|),\, \mathbb{E}(h_{l,0})$ from both sides)}} &\\
      &\leq \frac{\sigma_\delta^2}{\sigma_\delta^2 + \big(\rho(|\delta|)- \mathbb{E}(|\delta|)\big)^2} + \frac{\sigma_h^2}{\sigma_h^2 + \big(\rho(h_{l,0})- \mathbb{E}(h_{l,0})\big)^2}  {\text{(Cantelli's inequality)}}
\end{align*}
 where, $\sigma_\delta^2,\,\sigma_h^2$ are the variances of  $\delta$ and $h_{l,0}$ respectively. Note that as $\alpha$ increases, the risk gets larger as a greater value of $\alpha$ corresponds to a more risk-averse setting. Hence, the upper bound on $\mathbb{P}\{\text{MPC infeasible}\}$ gets smaller.
 
 We know that when $\alpha\rightarrow0$ (risk-neutral), $\rho(|\delta|)\rightarrow\mathbb{E}(|\delta|), \, \rho(h_{l,0}) \rightarrow \mathbb{E}(h_{l,0})$, and $\mathbb{P}\{\text{MPC infeasible}\}\leq1$. Similarly,  $\alpha\rightarrow1$ (risk-averse), $\rho(|\delta|)\rightarrow\max|\delta|, \, \rho(h_{l,0}) \rightarrow \max h_{l,0}$, and $\mathbb{P}\{\text{MPC infeasible}\}\rightarrow0$ (because $\mathbb{P}\{\delta_0 > \rho(|\delta|) \, \cup\, h_{l,0} > \rho(h_{l,0})\} \rightarrow 0$). Hence,~\eqref{eq:mpc_exact} is feasible at time $t+1$ if it is feasible at time $t$ with increasing probability as the confidence level $\alpha$ increases.
\end{proof}

\section{Proof of Proposition~\ref{proposition:task_completion}}
\begin{proof} (Adapted from~\cite{dixit2020risksensitive})
Consider the simple case of starting from $\boldsymbol{w}_{0}$ and reaching $\boldsymbol{w}_{1}$, i.e., when we have exactly two waypoints. We implement Algorithm \ref{alg:waypoint_alg} till $\psi_0 = 0$. Let $J_t^*$ be the cost of the MPC optimization~\eqref{eq:mpc_exact} at time $t$. The feasible solution to \eqref{eq:mpc_exact} at $t$ is given by the input sequence $\{(0, \eta_0^*),({K}_1^*, \eta_1^*),\dotsc , ({K}_{N-1}^*,\eta_{N-1}^*)\}$ and the state sequence $\{\psi_{0}^*, \psi_{1}^*, \dotsc ,\psi_{K}^*\}$. At time $t+1$, with confidence $\alpha$, the cost of the MPC optimization is 
$
J_{t+1}^* 
 \leq J_{t}^* - \lVert R\eta_0\rVert_1  - \psi_{0}^*.$
This is true because we know from Proposition \ref{proposition: rec_feasibility} that, with confidence $\alpha$, at time $t+ 1$, $\{(0,{K}_1^*\delta_0+ \eta_1^*),\dotsc , ({K}_{N-1}^*,\eta_{N-1}^*), (0, 0)\}$ is a feasible control input with $\psi(t+K|t+1) = 0$, i.e., $J_{t}^*$ will incur no additional cost from $(K_N, \eta_N|t+1)= (0, 0)$ and $\psi(t+K+1|t+1) = 0$. Since $J_{t+1}^* - J_{t}^* \leq - \lVert R\eta_0\rVert_1 - \psi_{0}^*$, the cost decreases by at least $1$ at each time step till the task is completed. Considering that $J_t^*$ is always positive and finite, it will take a finite number of steps to reach $J_k^* = 0, \, k\geq t$. Hence, the policy to take the system from $\boldsymbol{w}_{0}$ to $\boldsymbol{w}_{1}$ starting at time $t$ will be implemented in finite time, \textcolor{black}{in at most $\lceil J_{t}^*\rceil$ steps}, with confidence $\alpha$ \textcolor{black}{in each step  (as shown in Proposition~\ref{proposition: rec_feasibility}). Hence, the system will complete the task of traveling from $\boldsymbol{w}_{0}$ to $\boldsymbol{w}_{1}$ with confidence at least $\alpha^{\lceil J_{t}^*\rceil}$.} By induction, the system will reach $\boldsymbol{w}_K$ from $\boldsymbol{w}_0$ in finite time with confidence \textcolor{black}{$\alpha^{J_0^{K-1}}$}.
\end{proof}
\end{document}